\documentclass[amsmath,amssymb,aps,prd,twocolumn,superscriptaddress,10pt,nofootinbib]{revtex4-2}

\usepackage{graphicx}
\usepackage{dcolumn}
\usepackage{subfigure}
\usepackage{color}
\usepackage{url}
\usepackage{rotating}
\usepackage{xspace}
\usepackage[dvipsnames]{xcolor}
\usepackage[colorlinks,urlcolor=Blue,citecolor=Blue,linkcolor=Blue,pdfusetitle]{hyperref}
\usepackage{orcidlink}

\usepackage{amsmath}
\usepackage{amssymb}
\usepackage{graphicx}
\graphicspath{{figure_low_resolution/}}
\usepackage{dcolumn}
\usepackage{subfigure}
\usepackage{bm}
\usepackage{footnote}
\usepackage{url}
\usepackage{xspace}
\usepackage{xcolor}
\usepackage{makecell}
\usepackage{adjustbox}
\usepackage[normalem]{ulem}

\begin{document}

\title{Hierarchical search method for gravitational waves from stellar-mass binary black holes in noisy space-based detector data}

\author{Yao Fu\orcidlink{0009-0004-3032-9576}}
\affiliation{National Gravitation Laboratory, MOE Key Laboratory of Fundamental Physical Quantities Measurements, Department of Astronomy and School of Physics, Huazhong University of Science and Technology, Wuhan 430074, China}

\author{Yan Wang\orcidlink{0000-0001-8990-5700}}
\email{ywang12@hust.edu.cn}
\affiliation{National Gravitation Laboratory, MOE Key Laboratory of Fundamental Physical Quantities Measurements, Department of Astronomy and School of Physics, Huazhong University of Science and Technology, Wuhan 430074, China}

\author{Soumya D.~Mohanty\orcidlink{0000-0002-4651-6438}}
\email{soumya.mohanty@utrgv.edu}
\affiliation{Department of Physics and Astronomy, University of Texas Rio Grande Valley, 
Brownsville, Texas 78520, USA}
\affiliation{Department of Physics, IIT Hyderabad, Kandai, Telangana-502284, India}

\date{\today}

\begin{abstract}

Future space-based laser interferometric detectors, such as LISA, will be able to detect gravitational waves (GWs) generated during the inspiral phase of stellar-mass binary black holes (SmBBHs). These detections contain a wealth of important information concerning astrophysical formation channels and fundamental physics constraints. However, the detection and characterization of GWs from SmBBHs poses a formidable data analysis challenge, arising from the large number of wave cycles that make the search extremely sensitive to mismatches in signal and template parameters in a likelihood-based approach. This makes the search for the maximum of the likelihood function over the signal parameter space an extremely difficult task, with grid-based deterministic global optimization methods becoming computationally infeasible. We present a data analysis method that addresses this problem using both algorithmic innovations and hardware acceleration driven by Graphics Processing Units (GPUs). The method follows a hierarchical approach in which a semi-coherent $\mathcal{F}$-statistic is computed with different numbers of frequency domain partitions at different stages, with multiple particle swarm optimization (PSO) runs used in each stage for global optimization.  An important step in the method is the judicious partitioning of the parameter space at each stage to improve the convergence probability of PSO and avoid premature convergence to noise-induced secondary maxima in the semi-coherent $\mathcal{F}$-statistic. The hierarchy of stages confines the semi-coherent searches to progressively smaller parameter ranges, with the final stage performing a search for the global maximum of the fully-coherent $\mathcal{F}$-statistic.  We test our method on 2.5~years of a single LISA time delay interferometry (TDI) combination and find that for an injected SmBBH signal with a signal-to-noise ratio (SNR) between $\approx 11$ and $\approx 14$, the method can estimate (i) the chirp mass with a relative error of  $\lesssim 0.01\%$, (ii) the time of coalescence within $\approx 100$~sec, (iii) the sky location within $\approx 0.2$~${\rm deg}^2$, and (iv) orbital eccentricity at a fiducial signal frequency of $10$~mHz with a relative error of $\lesssim 1\%$.

\end{abstract}

\maketitle


\section{Introduction}
 Since the first direct detection of a gravitational wave (GW) signal from a stellar-mass binary black hole (SmBBH) merger, GW150914~\cite{abbott2016observation}, the twin LIGO observatories~\cite{aasi2015advanced} in coordinated observations with the Virgo~\cite{2015CQGra..32b4001A} and KAGRA~\cite{2012CQGra..29l4007S} detectors have recorded nearly $100$ events~\cite{abbott2016binary,abbott2019gwtc,abbott2021gwtc,abbott2021gwtc-3} during the first three observing runs (O1 to O3).
 Due to a lower frequency bound on the sensitive bandwidth of ground-based detectors arising from seismic noise, all observed signals have frequencies higher than $\approx 20$~Hz, which corresponds to the late inspiral, and merger phases of stellar-mass binary systems lasting for a $\sim 1$~min or less. Even with third-generation instruments in the future, such as the Einstein Telescope~\cite{ET2010} and Cosmic Explorer~\cite{CosmicExplorer2022}, ground-based detection of GW signals will ultimately be limited to $\gtrsim 1$~Hz due to gravity gradient noise~\cite{2011CQGra..28i4013H}. Observations of GW signals in the sub-Hz to millihertz range must necessarily use space-based detectors, projected to be operational in the next decade starting with the launch of the Laser Interferometer Space Antenna (LISA)~\cite{2017arXiv170200786A}. At around the same time, the proposed TianQin~\cite{2016CQGra..33c5010L,2021PTEP.2021eA107M} and Taiji~\cite{10.1093/nsr/nwx116,2020IJMPA..3550075R} (or one of them) may also join LISA to provide a network of space-based GW detectors that will have a significantly higher combined capability than any of the individual ones~\cite{2022PhRvD.106j2004Z,2021NatAs...5..881G,2023arXiv230716628T,2023arXiv230519714J}.

LISA will be sensitive to GW signals from a variety of sources in the millihertz range~\cite{2023LRR....26....2A,2022LRR....25....4A}, such as galactic compact binaries, extreme-mass-ratio inspirals (EMRIs), supermassive binary black hole mergers, SmBBHs, intermediate-mass binary black holes, and possibly the stochastic background from phase transitions and cosmic strings in the early Universe. Of particular interest to this paper is the fact that, complementary to ground-based detectors, LISA will be able to detect GWs in the early inspiral phase of SmBBHs in the chirp mass range $10-100~M_{\odot}$. It is expected that we will observe multiple SmBBHs~\cite{sesana2016prospects,kyutoku2016concise,wong2018expanding,moore2019stellar} throughout the mission lifetime of LISA and when combined with other proposed space-based GW detectors, such as TianQin~\cite{2016CQGra..33c5010L,2021PTEP.2021eA107M}, Taiji~\cite{10.1093/nsr/nwx116,2020IJMPA..3550075R}, and DECIGO \cite{2017JPhCS.840a2010S,2021PTEP.2021eA105K}, the number of detectable SmBBHs will be further increased~\cite{liu2020science}. 

Signals from SmBBHs will contain rich information, which will allow us to study their astrophysical formation channels~\cite{nishizawa2016elisa,breivik2016distinguishing,samsing2018black,gerosa2019multiband} and constrain fundamental physics theories~\cite{chamberlain2017theoretical,gnocchi2019bounding,tso2019optimizing,toubiana2020tests}. With a reasonable accuracy in the predicted time of coalescence of an SmBBH system that has a lifetime to merger of a few years, space-based detectors will be able to provide alerts to both ground-based GW detectors as well as electromagnetic observatories, enabling multi-wavelength and multi-messenger observations~\cite{sesana2016prospects,mcgee2020linking,caputo2020gravitational,vitale2016multiband} of such events.

However, in order to realize the rich scientific prospects outlined above we have to first overcome the major data analysis challenges~\cite{moore2019stellar} posed by the detection and parameter estimation of SmBBHs signals. Unlike ground-based detectors, SmBBH signals will persist in the sensitive frequency range of space-based detectors for several years, spanning more than $10^5$ wave cycles. This makes a matched filter-based search extremely sensitive to mismatches in the parameters of the signal and the templates, requiring an estimated  ($10^{30}-10^{40}$)~\cite{moore2019stellar} templates to cover the entire signal parameter space. In addition, each template, for 2.5 years of observations with a sampling interval of 10 seconds, will require $\approx 10^7$ samples, making the evaluation of each matched filter output computationally expensive. In fact, the computational cost associated with SmBBH signals is comparable to that of the more widely known problem of matching filter-based search for EMRI signals~\cite{cornish2011detection,chua2017augmented,babak2017science} and arises from essentially the same causes, namely, low amplitude signals lasting over many cycles requiring long integration times to accumulate sufficient signal-to-noise ratio (SNR).

Several studies have been conducted to assess the SmBBH data analysis challenge or propose exploratory methods to address it. For example, in ~\cite{liu2020science}, the Fisher information matrix was used to evaluate the parameter estimation accuracy of SmBBHs for LISA. As shown in ~\cite{ewing2021archival}, archival searches can uncover many low SNR GW signals in LISA data. 
In general, methods that have been proposed so far for SmBBH searches have strong underlying assumptions: In ~\cite{ewing2021archival,2024PhRvD.109f3029W}, the signal parameter search space is significantly restricted, while the performance of the methods in~\cite{buscicchio2021bayesian,toubiana2020parameter,2023arXiv230712244L} have only been studied for noiseless data.

In this paper, we present a data analysis method based on a hierarchical approach enabled by Particle Swarm Optimization (PSO)~\cite{kennedy1995particle,shi1998modified,mohanty2018swarm} that overcomes some of the above limitations and can search for SmBBH signals over a wide parameter range in the presence of noise, thus significantly advancing the state-of-the-art in this problem. 
PSO is a stochastic global optimization method that has been applied successfully to a wide range of GW data analysis problems~\cite{2017PhRvD..95l4030W,2020PhRvD.101h2001N,2021PhRvD.104b4023Z,2015ApJ...815..125W,2017PhRvL.118o1104W}, and proves to be highly effective in addressing the present problem too.
The hierarchical search is comprised of several stages of semi-coherent searches, inspired by the ones used extensively in continuous GW searches~\cite{riles2206searches,steltner2021einstein,steltner2023deep} with ground-based detectors, that are terminated by a fully coherent one.
An important feature of our method is the use of parameter space partitioning at each stage to avoid premature convergence to secondary noise-induced maxima of the semi-coherent search statistic and improve the convergence of PSO to its global maximum. In addition, we use multiple PSO runs at each stage and in each partition to further improve its performance.  

A hierarchical search has been investigated in~\cite{bandopadhyay2023lisa} for the case of noiseless data. Our method differs significantly in several aspects, such as the use of parameter space partitioning, maximization over all extrinsic parameters in the semi-coherent log-likelihood functions, and multiple PSO runs. We note that the problem of secondary maxima induced by noise is one of the major challenges in this search that does not appear in noiseless data.

The rest of the paper is organized as follows. Sec.~\ref{s2} describes in detail the search statistic. Sec.~\ref{s3} provides an overview of the hierarchical search method and its implementation in this paper. Sec.~\ref{s4} describes how we accelerate our code with the frequency-domain response of the detector and GPU-based parallelization. In Sec.~\ref{s5}, we present our main results on the performance of the method using simulated data realizations. Our conclusions and discussion of future improvements are presented in Sec.~\ref{s6}. Several technical details have been relegated to two appendices. Throughout this paper, we work in natural units in which $G = c = 1$.

\section{$\mathcal{F}$-statistic and Semi-Coherent Search}\label{s2}

In this section, we provide a brief overview of the Generalized Likelihood Ratio Test (GLRT) formalism, commonly called the fully-coherent $\mathcal{F}$-statistic~\cite{jaranowski1998data} in GW data analysis, along with our hierarchical search scheme based on the semi- and fully-coherent $\mathcal{F}$-statistic. 

\subsection{Fully-coherent $\mathcal{F}$-statistic}\label{s2.1}
 
The data from a space-based detector such as LISA consists of a set of time series $s = \{s^I(t)\}$, $t\in [0, T]$, where $I$ denotes a specific time-delay interferometry (TDI) combination~\cite{1999ApJ...527..814A,2005LRR.....8....4T} used to suppress laser frequency noise. Commonly used TDI combinations are the Michelson combinations $I\in \{X, Y, Z\}$ and combinations called $\{A, E, T\}$ that have mutually independent noise of instrumental origin~\cite{PhysRevD.66.122002}.

The detection of GW signals in noise requires deciding between the null hypothesis $H_0$, namely, the data $s$ does not contain a GW signal, and a set of alternative hypotheses, denoted as $H_1$, corresponding to the presence of an additive GW signal characterized by parameters $\Theta$. 
The most commonly used decision rule in searches for parameterized GW signals with theoretically known waveforms is the GLRT defined as~\cite{10.5555/151045}:
\begin{eqnarray}
\Lambda_G  &=& \max_\Theta \Lambda(\Theta|s)\;,\\
\Lambda(\Theta|s) &=& \ln\left(\frac{p(s | H_1;\Theta)}{p(s | H_0)}\right)\,,
\end{eqnarray}
where $\Lambda(\Theta|s)$ is called the log-likelihood ratio, and the joint probability density functions describing the
data under the hypotheses $H_0$ and $H_1$ (with signal parameters $\Theta$) are denoted by $p(s|H_0)$ and $p(s|H_1;\Theta)$, respectively. 

Assuming that the instrumental noise in each TDI combination is a realization of a stationary Gaussian stochastic process and that the TDI combinations chosen have mutually independent noise, 
\begin{equation}\label{eq2}
\Lambda(\Theta|s) = \sum_I \left[ \langle s^I, h^I(t;\Theta)\rangle_I - \frac{1}{2}\|h^I(t;\Theta)\|^2_I\right]\,. 
\end{equation}
Here, $h^I(\Theta)$ represents the GW signal, $\langle a,b \rangle_I$ is the noise-weighted inner product defined as~\cite{1992PhRvD..46.5236F}, 
\begin{equation}
\langle a,b\rangle_I = 4\,\mathrm{Re}\int_0^{+\infty} \frac{\widetilde{a}(f) \cdot \widetilde{b}^*(f)}{S_n^I(f)} df \,,
\label{eq:inner_product}
\end{equation}
where $\widetilde{a}(f)$ and $\widetilde{b}(f)$ are the Fourier transforms of continuous time functions $a(t)$ and $b(t)$, respectively, and $S_n^I(f)$ is the one-sided power spectral density (PSD) of the noise. The norm induced by the inner product is denoted as $\|a\|^2_I = \langle a, a \rangle_I$. 
For simplicity, in this work, we only employ the TDI X-combination for which the corresponding one-sided PSD is provided in~\cite{1999ApJ...527..814A,2017arXiv170200786A}. While adding data from other combinations is not fundamentally challenging, it increases the computational costs proportionally and exceeds the computational resources available for this study. As such, we drop the use of the TDI combination index $I$ from here on.

In this paper, we employ the restricted post-Newtonian waveform family described in Appendix~\ref{a1} that is characterized by a $10$-dimensional parameter space. The set of parameters is $\Theta = \{t_c, M_c, \delta_{\mu}, \theta, \phi, e_0, D_L, \psi, \iota, \phi_c\}$, which denote, starting from $t_c$, the time of coalescence, the chirp mass, the dimensionless mass difference, the sky location angles ($\theta = \pi/2 - \beta$, $\phi = \lambda$, where $\lambda$ and $\beta$ are the ecliptic longitude and ecliptic latitude, respectively), 
orbital eccentricity at a fiducial signal frequency of $10$~mHz, luminosity distance, polarization angle, the angle between the line of sight and the angular momentum direction, and the phase at the coalescence. We consider only the $(2,2)$ mode of the GW signal and leave the inclusion of higher-order modes arising from significant orbital eccentricity, spin, tidal effects, and other influences, to future work.

The GLRT involves maximizing the log-likelihood ratio defined in Eq.~\ref{eq2} over the space of signal parameters. The value of the global maximum serves as the detection statistic that is compared to a threshold, and if a detection occurs due to the crossing of the threshold, its location in parameter space provides the Maximum Likelihood Estimate (MLE) of the parameters of the true GW signal. One of the main challenges across many GW data analysis problems, including the SmBBH search, is the extremely high computational cost associated with deterministic methods for solving the global optimization over $\Theta$ required in the GLRT.

It is straightforward to show that the response $h(t; \Theta)$ can be written as~\cite{jaranowski1998data,crowder2007solution}
\begin{equation}\label{eq:response}
h(t, \Theta) = \sum_{i=1}^{4} a_i A^i(t; \Theta_I) \,,
\end{equation}
where $\Theta_E = \{a_i\}$, which are time-independent reparameterizations of $\{D_L, \psi, \iota, \phi_c\}\subset \Theta$, comprise the set of extrinsic parameters, while $\Theta_I = \{t_c, M_c, \delta_{\mu}, \theta, \phi, e_0\}$ form the set of intrinsic parameters. 
This allows $\Lambda_G$ to be expressed as 
\begin{equation}\label{eq:exinparam} 
\Lambda_G = \max_{\Theta_I}\left(\max_{\Theta_E}\Lambda(\Theta|s)\right) \,.
\end{equation}
One can analytically maximize $\Lambda(\Theta|s)$ over $\Theta_E$ to get the fully-coherent $\mathcal{F}$-statistic~\cite{crowder2007solution,jaranowski1998data}, 
\begin{equation}
\mathcal{F}(\Theta_I|s) = \max_{\Theta_E} \Lambda(\Theta|s) = \frac{1}{2} N^T(\Theta_I) M^{-1} N(\Theta_I) \,,
\label{eq:coherent_F_stat}
\end{equation}
where,  
\begin{eqnarray}
N^i(\Theta_I) & = & \langle s|A^i(\Theta_I)\rangle \,, \label{aandMN}\\
M^{ij} & = & \langle A^i | A^j\rangle \,, \label{aandMN2} \\
a & = & N^TM^{-1} \,. \label{aandMN3}
\end{eqnarray}
The dimensional reduction above from $10$ to $6$ significantly reduces the difficulty of the global optimization problem. Once the MLE estimates of $\Theta_I$ are obtained, the corresponding MLE estimates of $\Theta_E$ are obtained as shown in Appendix~\ref{a2}. 

As mentioned earlier, the global optimization of the fully-coherent $\mathcal{F}$-statistic over the full astrophysical search range in $\Theta_I$ space, called a fully-coherent search, is computationally infeasible if carried out deterministically using a grid of points in this space. Stochastic optimization methods are also likely to fail due to the needle-in-the-haystack problem posed by the extremely small footprint of high values around the global maximum compared to the required search range. A practical solution is to use a hierarchical approach in which the search space is first narrowed down to some promising regions in $\Theta_I$ space, making follow up fully-coherent searches in each region computationally feasible. We follow this approach using a semi-coherent version of the $\mathcal{F}$-statistic as described below.
Henceforth, in the following, we use the term search space to refer to the multi-dimensional volume being searched in any one run of PSO while search range refers to the length of the interval along one or more dimensions of this volume.

\subsection{Semi-coherent $\mathcal{F}$-statistic}\label{s2.2}

The semi-coherent step is inspired by the principal idea behind current continuous wave searches using ground-based GW detectors~\cite{riles2206searches,steltner2021einstein,steltner2023deep}. This is the partitioning of the time domain GW strain data from a detector into smaller segments that are individually subjected to fully coherent searches. Instead of finding just the global optimum in each search, a set of significant locations in $\Theta_I$ are found. Different schemes have been proposed and implemented~\cite{2000PhRvD..61h2001B,2005PhRvD..72d2004C,2004PhRvD..70h2001K,Antonucci_2008} for combining these locations into candidates for follow up with fully-coherent searches using longer segment lengths. It should be noted that there are alternative approaches~\cite{2008PhRvD..77h2001D}  possible in how the initial step of the hierarchy is implemented that do not use the $\mathcal{F}$-statistic at all. To limit computational costs, the threshold in the initial step of the above types of hierarchical approaches must be set such that the number of false alarms that need to be followed up becomes manageable. As such, semi-coherent searches generally incur a trade-off between available computational resources and loss in detection sensitivity.

The primary motivation for using time-segmentation in ground-based continuous wave searches is the high cost of computing Discrete Fourier Transforms (DFTs) due to the large sampling rate of the data. However, this is not an issue for data from space-based detectors since the number of data samples from the entire mission lifetime of LISA would be $\sim 10^8$, which is just about the size of a ten hours segment from a single LIGO detector sampled at $2048$~Hz. Instead, for space-based detectors, the hierarchical approach can be implemented using frequency domain partitioning and 
this is the approach we follow here.

Let $\widetilde{\mathcal{F}}(\Theta_I|s_l)$ denote the $\mathcal{F}$-statistic function computed using data $s_l$ that is band-limited to $[f_l^{(0)}, f_l^{(1)}]$ and let 
\begin{eqnarray}\label{eq:semi-cohF}
\widehat{\mathcal{F}}(\Theta_I) &=& \frac{1}{N_{\text{seg}}}\sum_{l=1}^{N_{\text{seg}}} \widetilde{\mathcal{F}}_l(\Theta_I) \;,\nonumber\\
& = & \frac{1}{N_{\text{seg}}}\sum_{l=1}^{N_{\rm seg}} \max_{\Theta_E} \Lambda(\Theta|s_l) 
\end{eqnarray}
define the semi-coherent $\mathcal{F}$-statistic, in which $N_{\rm seg}$ is the number of segments. 
Since the inner product can be computed in the frequency domain following Eq.~\ref{eq:inner_product}, $\Lambda(\Theta_I|s_l)$ is obtained by simply redefining the inner product for segment $l$ as,
\begin{equation}
    \langle s,h \rangle_l = 4 \, \text{Re} \int_{f_l^{(0)}}^{f_l^{(1)}} \frac{\widetilde{s}(f) \cdot \widetilde{h}^*(f)}{S_n(f)} df \,. 
\end{equation}

For continuous GW searches in ground-based detectors, where the signal is nearly monochromatic, the SNR and the number of wave cycles are nearly evenly distributed across time and, hence, the semi-coherent search can use uniform partitioning of the data in the time domain. In contrast, this is not true for SmBBH signals since the GW power is neither uniformly distributed in time nor in frequency since these are evolving sources. One possible approach is to determine the frequency-domain segment boundaries based on the SNR in each segment. 
However, as mentioned in~\cite{bandopadhyay2023lisa}, this partitioning scheme introduces a dependence of the segment boundaries on the template parameters, necessitating a recalculation of the segment limits for each evaluation of $\widehat{F}$. This can increase program complexity and computation time.
Therefore, we opt to compute $f_l^{(1)}$ and $f_l^{(0)}$ by fixing the number of segments for all templates but keeping  $f_1^{(0)}$ and $f_{N_{\rm seg}}^{(1)}$ dependent on some of the intrinsic parameters. In our implementation,  these limits are set to be the lowest and highest instantaneous frequencies occurring during the observation period in the template associated with $\Theta_I$.

\begin{figure*}[htbt!]
\includegraphics[width=1\textwidth]{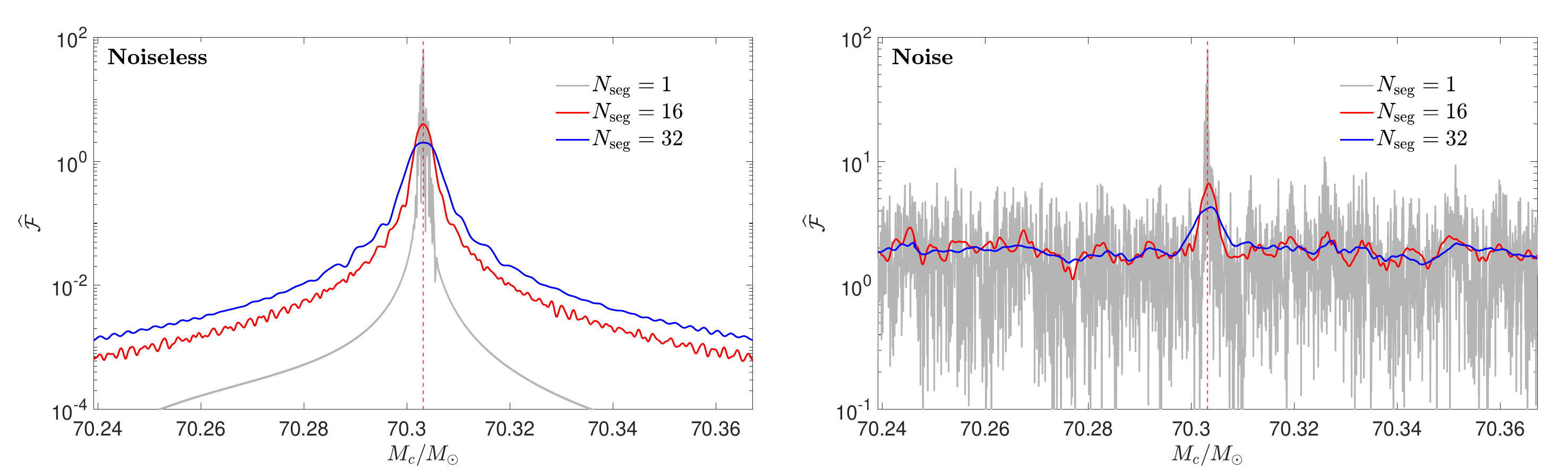}
  \caption{
  Comparison of the semi-coherent $\mathcal{F}$-statistics ($N_{\text{seg}}>1$) with the fully-coherent $\mathcal{F}$-statistic ($N_{\text{seg}}=1$) under different conditions. The true value of the chirp mass, $M_{c} = 70.3032M_{\odot}$, is marked with a red dashed line,  while the other parameters are fixed at their true values, and the SNR is 11.3.
  In the left panel, one can see that, under noiseless condition,  the peaks for the semi-coherent $\mathcal{F}$-statistics are significantly broadened compared to the fully-coherent $\mathcal{F}$-statistic, which can mitigate the difficulty of the search for the global maximum. However, under noisy conditions, as shown in the right panel, as $N_{\text{seg}}$ increases, the peaks become less distinct relative to the fluctuations caused by noise. 
  }
  \label{fig:1_123}
\end{figure*}

Unlike the fully coherent case, the calculation of the semi-coherent $\widehat{\mathcal{F}}(\Theta_I)$ allows different values of the extrinsic parameters $\Theta_E$ for each segment. This reduces the sensitivity of the detection statistic to the intrinsic parameters, and the SNR in each segment is now lowered. However, the profile of the $\mathcal{F}$-statistic is significantly broadened around the global maximum of the fully-coherent $\mathcal{F}$-statistic, which reduces the difficulty of the search considerably. Fig.~\ref{fig:1_123} illustrates, in both the absence and presence of noise, the profiles of the semi-coherent $\mathcal{F}$-statistic with different $N_{\rm seg}$ as a function of the chirp mass $M_{c}$. Here, other parameters are set to their true values. Note that $N_{\rm seg} = 1$ is equivalent to the case of the fully-coherent $\mathcal{F}$-statistic. 

Fig.~\ref{fig:1_123}(a) shows how, in the absence of noise, the semi-coherent method expands the width of the peak and absorbs the sharp secondary peaks near the true value of $M_{c}$ into the main peak. This makes it easier to search for the peak. However, as shown in Fig.~\ref{fig:1_123}(b), the peak also becomes less prominent compared to the fluctuations away from the peak caused by noise. 
Thus, adopting an appropriate $N_{\rm seg}$ according to the type of the GW signal is a key issue in a semi-coherent search.

\section{Hierarchical search pipeline}\label{s3}

In this section, we provide a brief overview of Particle Swarm Optimization (PSO) and discuss the specific implementation used in this work. This is followed by a description of the search pipeline that uses PSO in multiple
stages of a hierarchical search pipeline.

\subsection{PSO}\label{s3.1}
PSO is an iterative stochastic search method for the global maximum of a function $f(x)$, $x \in \mathbb{R}^N$, called the fitness function, over a specified compact subset $D\subset \mathbb{R}^N$ called the search space. [In our case, the fitness function is the semi-coherent $\mathcal{F}$-statistic (c.f., Eq.~\ref{eq:semi-cohF}) and the search space is the normalized space of intrinsic parameters $\Theta_I$.] In each iteration, the fitness is sampled at multiple locations and these are updated for the next iteration based on the sampled fitness values. The sampled locations, which are typically fixed in number, are called particles in PSO, with the rules used for updating their positions, called the dynamical equations, stated below. The collection of particles is called a swarm.  

Let $x_{i,j}[k]$ denote component $j$ of the position vector $x_i$ of particle $i\leq N_{\rm part}$ at iteration $k$. Attached to each particle is a displacement vector $v_i[k]$ (called velocity in PSO terminology) with component $j$ denoted as $v_{i,j}[k]$. The position and velocity are  updated as,
\begin{equation}\label{eq:pso_position_update}
x_{i,j}[k+1] = x_{i,j}[k] + \min(v_{i,j}[k+1], v_{\rm max}) \,,
\end{equation}
\begin{equation}\label{eq:pso_velocity_update}
\begin{aligned}
v_{i,j}[k+1] &= w[k] v_{i,j}[k] + c_1 r_1 (p_{i,j}[k] - x_{i,j}[k]) \\
&+ c_2 r_2 (l_{i,j}[k] - x_{i,j}[k])  \,,
\end{aligned}
\end{equation}
where $v_{\rm max}$ limits the maximum step size in any dimension, $w[k]$ denotes a deterministically decaying number called the inertia weight, $c_{1,2}$ are called acceleration constants, $r_{1,2}\sim U([0,1])$ are independent random variables with uniform distribution over $[0,1]$, $p_i[k]$ (personal best) is the location with the best fitness in the history of particle $i$, and $l_i[k]$ (local best) is the best location found in the history of its local neighborhood. (When the local neighborhood includes all the other particles in the swarm,  $l_i[k]$ becomes the global best, $g[k]$.) An example of a neighborhood definition is the ring topology~\cite{4223164} in which particle indices are arranged on a circle and a subset of consecutive indices constitutes a neighborhood.

Each of the three terms in Eq.~\ref{eq:pso_velocity_update} has a special role. The term with the inertial weight, called the inertia term, causes the particle to move past its current location irrespective of its fitness value. This allows the particle to escape local maxima and contributes to the exploration of the search space. The second term, called the cognitive term, that depends on the personal best can be interpreted as a random force that tries to attract the particle back to a good location that it has already found. The last term, called the social term, that depends on the local best is also a random force that seeks to pull the particle to a good location found by its neighbors. The cognitive and social terms promote exploration around previously found good solutions, a stage of the search called exploitation, but the randomness in these terms prevents too quick a convergence. Thus, the dynamical equations create a trade-off between the exploration and exploitation behavior of the swarm, with the former dominating in the beginning. The use of a local instead of a global best extends the exploration phase by slowing down information flow within the swarm. The inertia weight $w[k]$ is typically allowed to decay linearly with iterations, providing yet another control mechanism for the shift in swarm behavior from exploration to exploitation.

Most PSO implementations work best when the search space $D$ is a hypercube, with component $j$ of $x$  limited to $[a_j, b_j]$. The iterations are commonly initiated with random locations $x_{i,j}[0]\sim U([a_j,b_j])$ and velocities $v_{i,j}\sim U[a_j - x_{i,j}, b_j - x_{i,j}]$. While these initial conditions guarantee that particles will be localized within $D$ in the first position update, it is possible for them to escape $D$ in subsequent iterations. Therefore, boundary conditions need to be specified to handle such particles. Among the many boundary conditions proposed in the literature~\cite{engelbercht2005fundamentals}, we choose the let-them-fly condition in which nothing is done to the particles except for setting their fitness to $-\infty$ (in a maximization problem) when they escape $D$. Similarly, among the diverse termination conditions proposed for PSO, we pick the simplest one in which termination happens once a specified number of iterations is completed. The final solution for the maximization problem is the global best location and the corresponding fitness value. Although this criterion leads to an excess number of fitness evaluations, it allows load balancing in a parallel implementation of multiple PSO runs and leads to a more efficient utilization of computing resources.

Like most other stochastic heuristics for global optimization, there is no guarantee of convergence (even asymptotically) to the global maximum in PSO 
but it can be tuned to provide an acceptable probability, $P_{\rm success}$, of successful convergence to a given region containing the global maximum. 
This probability can be boosted exponentially by running $N_{\rm runs}$ independent runs of PSO, with statistically independent pseudorandom streams for $r_{1,2}$ and initialization, and picking the best solution across all the runs. The resulting probability of success becomes $1 - (1-P_{\rm success})^{N_{\rm runs}}$, which increases rapidly with $N_{\rm runs}$ even for moderate $P_{\rm success}$.

Much of the overview provided above is common to several applications of PSO in GW data analysis. In fact, we have adopted the same settings for most of the parameters above as given in~\cite{mohanty2018swarm} except for $N_{\rm part} = 128$, which is significantly higher than the usual choice of $\approx 40$ particles, and different $N_{\rm runs}$ at different stages of the hierarchical search. The large number of particles used here follows from the specific implementation of GPU acceleration used in our code, which requires $N_{\rm part}$ to be a power of $2$, and the fact that lower $N_{\rm part}$ values of $32$ and $64$ did not yield a high probability of successful convergence. As described below, the number of PSO iterations used in the termination criterion varies for the different stages of the hierarchy. The remaining PSO parameters are set to widely used values~\cite{4223164}: $c_1 = c_2 = 2.0$ and $v_{\rm max} = 0.5$, and because the search range is normalized for PSO, these parameters are dimensionless and the magnitude is independent of the actual search range.

\subsection{Hierarchical search}\label{s3.2}

\begin{figure*}[htbt!]
\includegraphics[width=1\textwidth]{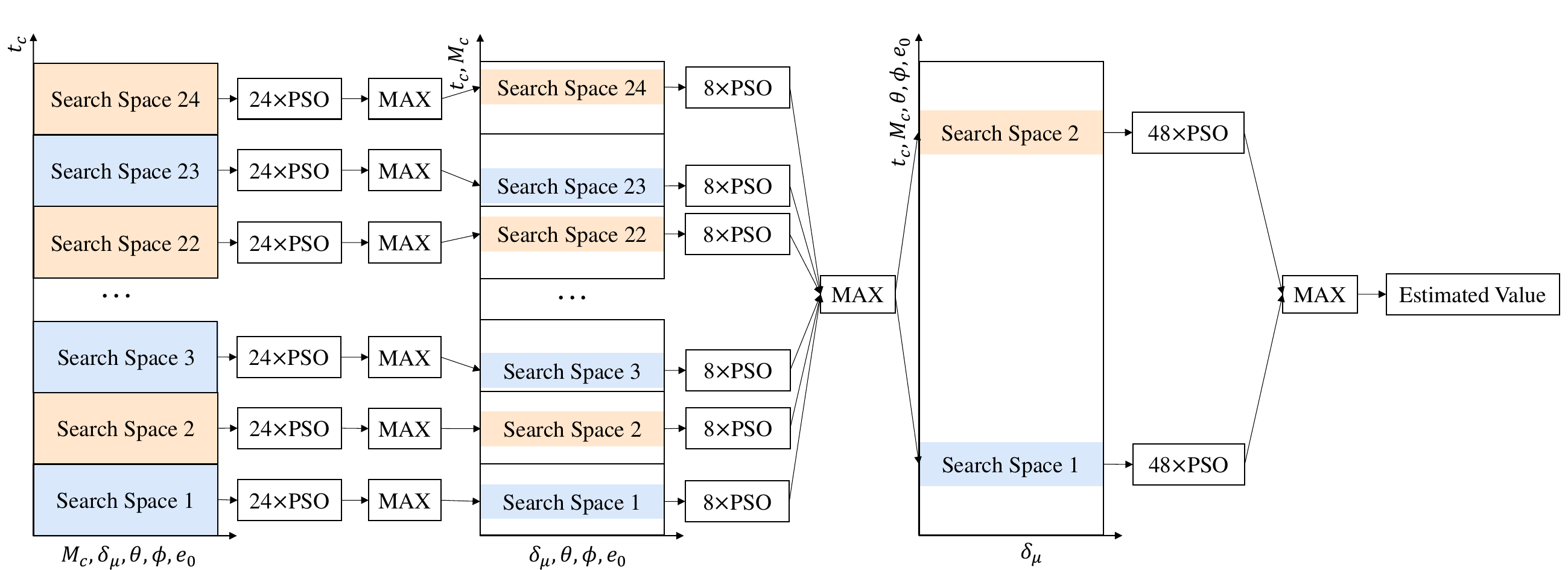}
\caption{\label{fig:2_program} Flowchart of the hierarchical stages. Each stage is shown by a stack of colored boxes (the boxes are colored with alternating colors for visual clarity). Each stack has two axes: the horizontal and vertical axes are labeled by parameters for which the search space is not subdivided or divided, respectively. Each colored box schematically represents one partition of the search space for the latter set of parameters. In stage--1, starting at the left, the search range of $t_c$ is uniformly divided into 24 contiguous intervals. In stage--2, based on the results of stage--1, the search ranges of both $t_c$ and $M_c$ are narrowed down, while the search ranges of other parameters remain unchanged. In stage--3, the search ranges of all parameters except $\delta_\mu$ are narrowed down. Each colored box is followed by a box showing the number of PSO runs used in that restricted search space. Each box labeled MAX represents the operation of selecting the best solution out of the PSO runs connected to it. Finally, in stage--3, the two search spaces differ in the range of $\theta$, with the search ranges over all the other parameters kept identical.}
\end{figure*}

Despite the significant broadening of the peak in the fitness function, as shown in Fig.~\ref{fig:1_123}, it is still much smaller in width than the full search range used in our analysis. In addition, the semi-coherent fitness function is very rugged with multiple local maxima induced by noise in the data. 
Therefore, PSO would have a low probability of success if it is used over the full search space.
To cope with this problem, we divide the search space into smaller regions along certain dimensions, search over the regions independently, and propagate their results in a hierarchical manner through multiple stages in order to progressively narrow down the search around the global maximum of the semi-coherent $\mathcal{F}$-statistic. The details of this hierarchical pipeline are provided below and illustrated schematically in Fig.~\ref{fig:2_program}.
\textbf{Stage--1:} The search range of the parameter $t_c$ is divided into $24$ equally wide and contiguous intervals of length $\approx 0.02$ yr. 
In each interval, $24$ independent PSO runs are carried out to search for the global maximum of the semi-coherent $\mathcal{F}$-statistic using $N_{\rm seg}=16$ segments. The best solution among the $24$ runs in each region is then recorded, producing a total of $24$ candidate locations for the global maximum in the full parameter space. The number of iterations in this step is $8000$ for each PSO run.

\textbf{Stage--$2$:} In this stage, the $24$ candidate locations obtained in stage 1 form the centers for new search spaces that are restricted in both $t_c$ and $M_c$. Based on our experience with different settings, we set the size of each restricted range to $0.01$~yr along $t_c$ and $0.6$~$M_{\odot}$ along $M_c$, while keeping the search ranges for all other parameters the same as in stage~$1$. Therefore, the size of the search spaces in the Stage 2 is two orders of magnitude smaller than in the Stage 1. Each of the $24$ regions is now searched with $8$ PSO runs and the single best solution from all these $24\times 8 = 192$ runs is propagated to the next stage. The number of PSO iterations and $N_{\text{seg}}$ for semi-coherent $\mathcal{F}$-statistic remain the same as in stage~$1$. The single solution provides a candidate location in a search space that is restricted in $t_c$ and $M_c$ but unrestricted in all other parameters.

\textbf{Stage--3:} In this stage, the single solution obtained in the previous stage forms the center of narrower search ranges with $0.004$~yr for $t_c$, $0.3$~$M_\odot$ for $M_c$, $0.4$~rad for $\theta$, $0.4$~rad for $\phi$, and $0.006$ for $e_0$. As for $\delta_\mu$, due to the relative insensitivity of the semi-coherent $\mathcal{F}$-statistic to this parameter, we continue to maintain its original search range. Besides restricting the search ranges, we also create a search space centered on $\pi -\theta$, due to a degeneracy of the $\mathcal{F}$-statistic in $\theta$. Therefore, the size of the search spaces at this time is seven orders of magnitude smaller than in the Stage 1. Each of the two search spaces is now subjected to $48$ PSO runs with 15,000 iterations per run. At this point, $N_{\text{seg}}$ is set to 1, which means that this stage uses the fully-coherent $\mathcal{F}$-statistic. The best solution across the two search spaces provides the final estimate of the intrinsic parameters that are then used to obtain the corresponding estimates of the extrinsic parameters following the expressions given in Appendix~\ref{a2}.

While we restrict the search ranges for only $t_c$ and $M_c$ in the first two stages, it is conceivable to have a more general scheme in which the search ranges of additional parameters are restricted. However, our experience shows that this is redundant in the first two stages because the semi-coherent $\mathcal{F}$-statistic is most sensitive to these two parameters and relatively insensitive to the others. Moreover, subdividing the ranges of the parameters increases the computational cost as the number of such parameters goes up. Thus, subdividing the ranges of only $t_c$ and $M_c$ offers a compromise between effectively locating the interesting search space for the latter stages and the overall computational cost of the search.
The number of search spaces in the different stages and the number of PSO runs per search space have been obtained empirically based on their effect on the overall performance of the search method. With additional computing resources, these numbers can be increased and the performance can be improved.

It is worth noting that we do not reduce the number of search spaces when going from stage--1 to stage--2. This is because, in the stage-1 search, there is a high probability that secondary maxima induced by noise in search spaces far from the one containing the true parameter have larger values. Therefore, confining the stage-2 search to a subset of the stage-1 search ranges incurs the risk of prematurely converging to secondary maxima. It is preferable to further refine the solution in each search range, as done in stage--2, before deciding to pick the best solution.

\section{Efficient computation of the semi-coherent $\mathcal{F}$ statistic}\label{s4}

The hierarchical stages described above involve a large number of PSO runs over independent search spaces, with each run requiring a large number of iterations. To mitigate the computational costs involved, we use an efficient calculation of the frequency domain TDI response to a plane GW and implement GPU acceleration of parallelizable steps.

The computationally most expensive step in PSO is the evaluation of the semi-coherent $\mathcal{F}$-statistic, $\widehat{F}$, for each particle. Calculating $\widehat{F}$ involves computing the noise-weighted inner product between the templates and the data, which requires Fourier transforms and consumes a significant amount of time.

A TDI combination is constructed as a sum of time-delayed single-arm responses. 
The individual contribution of the GW-induced single-arm response can be written in the following form~\cite{2005PhRvD..71b2001V,marsat2018fourier}:
\begin{equation}
x(t) = F(t)h(t + d(t)) \,. 
\end{equation}
Here, $F(t)$ is the time-domain modulation associated with the rotating antenna pattern of the arm and precessional effects in the source orbit evolution. 
Note that the former depends on the orbit of the detector. For LISA, the orbit used in this study is an analytic approximation, with accuracy extended to second-order eccentricity~\cite{2003PhRvD..67b2001C}. While this provides convenience and efficiency, it is worth noting that our work does not depend on it. Numerical orbits with higher precision can also be used, and the evaluation of the detector antenna pattern at specific times requires numerical interpolation using appropriate orders of Chebyshev polynomials, as demonstrated in~\cite{2021PhRvD.103j3026Z}.
$d(t)$ represents the time delay due to the orbital motion of the detector. 
A GW chirp signal, such as that from an SmBBH inspiral, has a well-defined instantaneous frequency and its Fourier transform can be expressed as~\cite{2005PhRvD..71b2001V,PhysRevD.69.082003,1975GReGr...6..439E}: 
\begin{equation}
\widetilde{h}(f) = A(f) e^{i\Phi(f)} \,,
\end{equation}
where $A(f)$ and $\Phi(f)$ can be obtained using the stationary phase approximation. 
According to \cite{marsat2018fourier}, for sources where the $T_f$ is much smaller than the timescale of the detector modulation, the Fourier transform of $x(t)$ is well approximated by:
\begin{equation}\label{eq:freDres}
\widetilde{x}(f) = T(f)\widetilde{h}(f) \,,
\end{equation}
where 
\begin{widetext}
\begin{equation}
T(f) = \sum_{k\geq 0} \frac{(-i)^k}{k!} (T_{Ak})^k F_{T_{f,\epsilon}} \left[ \frac{d^k}{d\tau^k} \left( \frac{F(\tau-d(\tau))}{1+\dot{d}(\tau)} e^{2i\pi fd(\tau)(1-\dot{d}(\tau))} \right) \right](t_f) \,.
\end{equation}
\end{widetext}
Here, 
\begin{equation}
   T_{A1} = \frac{7}{6} \frac{1}{2\pi f} \,, 
\end{equation}
\begin{equation}
   t_f = \frac{1}{2\pi} \frac{d\Phi(f)}{df} \,, 
\end{equation}
\begin{equation}
    T_f^2 = \frac{1}{4\pi} \left| \frac{d^2\Phi}{df^2} \right| \,, 
\end{equation}
\begin{equation}
    \epsilon = -\text{sgn}\left(\frac{d^2\Phi}{df^2}\right) = -1 \,, 
\end{equation}
\begin{equation}
    F_{T_{f,\epsilon}}^N(F)(t) \equiv \frac{1}{2} \sum_{k=0}^N a_{N,k}^{\epsilon} \left( F(t+kT_f) + F(t-kT_f) \right) \,,
\end{equation}
where $a_{N,k}^{\epsilon}$ are defined in Appendix 2 of \cite{marsat2018fourier}.
In this work, all the SmBBHs that we consider have $T_f$ less than $10^6$~s, while the modulation timescale of LISA is $\approx 3\times 10^{7}$~s, which is consistent with the approximation used in~\cite{marsat2018fourier} for frequency-domain response in Eq.~\ref{eq:freDres}.
Thus, even if we set $N$ to zero, and neglect the higher-order terms of $d(t)$ in deriving the frequency-domain response, the match with the approximation-free frequency-domain signal still exceeds 0.999. Using the frequency-domain response in Eq.~\ref{eq:freDres} to compute $\widehat{F}$ not only simplifies the computation since time-domain calculation of the responses are not required but also allows deep parallelization of the calculation.

In the hierarchical method presented in Sec.~\ref{s3}, there are several nested parallelizable steps.
First, each PSO run can be executed independently. Secondly, within each PSO run, the computation of the semi-coherent $\mathcal{F}$-statistic for each particle can be carried out in parallel. Finally, in the frequency-domain response, the computations for each frequency can be parallelized. We implement these layers of parallelization using a nested set of hardware elements, starting with nodes of a cluster at the top level, CPU cores at the intermediate level, and GPUs at the deepest level. At present, our parallelization scheme allows the code implementing the hierarchical search described in Sec.~\ref{s3} to perform a complete analysis of a single data realization in $\approx 2.5$~days using $8$ NVIDIA A100 GPUs attached to a single compute node with $48$ CPU cores. 

In the current state of development of data analysis methods for SmBBHs, the comparison of computational costs is quite difficult because different methods approach the problem with very different objectives (e.g., archival search ~\cite{ewing2021archival,2024PhRvD.109f3029W}, search in noise free data ~\cite{buscicchio2021bayesian,toubiana2020parameter,2023arXiv230712244L,bandopadhyay2023lisa}, etc.) and the computation time and hardware used are generally not reported in adequate detail. In addition, when comparing computational costs one must account for the SNR of the signal being injected since it affects the performance of the search.

\section{Results}\label{s5}

In this section, we discuss the results obtained by applying our pipeline to different sets of simulated data.

\subsection{Noiseless data}\label{s5.noiseless}

First, we validate the hierarchical search method on noiseless data. The values of the parameters and corresponding search ranges for the simulated GW signals are given in Table~\ref{tab:GW_param_no_noise}. 
Since there is no noise in the data, we can adjust some of the configuration parameters to increase the execution speed. For example, we set $N_{\text{seg}} = 128$ in both stage--1 and stage--2, which significantly reduces the search complexity by broadening the global maximum of the semi-coherent $\mathcal{F}$-statistic while not incurring any risk of noise-induced secondary maxima. At the same time, we reduce the number of PSO iterations to 2000 for the first two stages and 5000 for the stage--3. Finally, the number of PSO runs per search space was reduced substantially. These adjustments reduced the runtime of the program by a factor of $10$ compared to the fiducial settings given in Sec.~\ref{s3}. 
\begin{table}[htbp]
    \centering
    \caption{The parameters used in the search of noiseless data and their search ranges. The signal amplitude is chosen such that the SNR relative to the instrumental noise has the value shown here. $D_L$, $\iota$, $\psi$, and $\phi_c$ are extrinsic parameters, whose estimated values are obtained from the relationship given in the Appendix~\ref{a2}. Hence, they do not have an associated search range.}
    \label{tab:GW_param_no_noise}
    \begin{ruledtabular}
    \begin{tabular}{ccc}
    
    Parameter      & {True value}   & {Search range}   \\ \hline
    $t_c/{\rm yr}$       & 3.5000          &[3.3,\,3.8]       \\
    $M_c/M_{\odot}$ & 70.3032         &[20,\,80]           \\
    $\delta_\mu$   & 0.1585          &[0,\,0.9]           \\
    $\theta/{\rm rad}$   & 0.7854          &[0,\,$\pi$]         \\
    $\phi/{\rm rad}$     & 2.1216          &[0,\,$2\pi$]       \\
    $e_0(f_{\rm gw}=0.01\,\text{Hz})$ &0.0100&[0,\,0.05]          \\
    $D_L/{\rm Mpc}$      & 920.9084        & /                \\
    $\iota/{\rm rad}$    & 0.5329          & /                \\
    $\psi/{\rm rad}$     & 1.2501          & /                \\
    $\phi_c/{\rm rad}$   & 0.8419          & /                \\
    ${\rm SNR}$          & 11.3137         &/                 \\

    \end{tabular}
    \end{ruledtabular}
\end{table}

Table~\ref{tab:no_noise} presents a comparison of the true values with the estimated ones for noiseless data. It can be seen that, except for $\delta_\mu$, the intrinsic parameters are recovered with very small relative errors.  Part of the error in $\delta_\mu$ can be attributed to degeneracies between $\delta_\mu$ and several other parameters in the $\mathcal{F}$-statistic. To test this hypothesis, we reran the analysis fixing $\delta_\mu$ to its true value. The corresponding results displayed in Table~\ref{tab:no_noise} show that errors in almost all the parameters are reduced with some of the reductions being significant. This shows that allowing $\delta_\mu$ to be a free parameter causes errors in other parameters to be compensated by an error in $\delta_\mu$. 
Table~\ref{tab:no_noise} also contains the estimated values of the extrinsic parameters calculated by the procedure in Appendix~\ref{a2}. It can be seen that the relative errors in these parameters are orders of magnitude larger than those in the intrinsic parameters. We believe that this is due to some degree of degeneracy within the extrinsic parameters~\cite{2023CoTPh..75g5402W}. In the future, using more accurate waveforms that include higher-order modes should improve the accuracy of estimation for the extrinsic parameters~\cite{PhysRevD.99.124005,PhysRevD.108.064046}.

Thus, the analysis of noiseless data validates our code and allows us to proceed to the analysis of realistic noisy data.

\begin{table*}[htbp]
    \centering
    \caption{Results of the search in noiseless data. It can be seen that the intrinsic parameters, except $\delta_\mu$, are close to their true values. However, there is a considerable error in the extrinsic parameters. This is due to the problem of degeneracy in the GW waveform. Even in a noise-free scenario, obtaining a highly accurate value for these parameters remains challenging. The results of the search when $\delta_\mu$ is fixed at the true value are also given here, and it can be seen that the error is basically further reduced.}
    \label{tab:no_noise}
    \begin{ruledtabular}
    \begin{tabular}{cccccc}
    
    Parameter    & {True value}  &{Estimated value}  &{Relative error}  &{Estimated value~($\delta_\mu$ fixed)}  &{Relative error}\\ \hline
$t_c/\rm yr$       &3.5000000 & 3.5000029 & 0.0000836\% &3.4999994&0.0000172\%\\
$M_c/M_{\odot}$ &70.30316 & 70.30662 & 0.00492\% &70.30319&0.0000436\%\\
$\delta_\mu$   &0.1585 & 0.4784 & 201\% &/&/\\
$\theta/\rm rad$   &0.78540 & 0.78554 & 0.0174\% &0.78535&0.00621\%\\
$\phi/\rm rad$     &2.12160 & 2.12162 & 0.00128\% &2.1214&0.00788\%\\
$e_0(f_{\rm gw}=0.01\text{Hz})$&0.01000000 & 0.00998489 & 0.151\%&0.00999905&0.00947\% \\
$D_L/\rm Mpc$      &920.9084 & 880.9283 & 4.34\% &895.3753&2.77\%\\
$\iota/\rm rad$    &0.5329 & 0.6088 & 14.2\% &0.5826&9.33\%\\
$\psi/\rm rad$     &1.2501 & 1.2846 & 2.76\% &1.2831&2.64\%\\
$\phi_c/\rm rad$   &0.8419 & 0.2348 & 72.1\% &0.7808&7.26\%\\

    \end{tabular}
    \end{ruledtabular}
\end{table*}

\subsection{Noisy data}\label{s5.noise}

Within the analysis of noisy data, we considered three separate cases, labeled as cases A, B, and C. 
Table~\ref{tab:CaseABC} provides the values of the true signal parameters in the three cases. The search ranges for all three cases are identical to the ones for the noiseless case listed in Table~\ref{tab:GW_param_no_noise}. In the following, we discuss the results from these cases in sequence. 
\begin{table}[htbp]
    \centering
    \caption{The parameters in Case A, Case B, and Case C and their respective search ranges.}
    \label{tab:CaseABC}
    \begin{ruledtabular}
    \begin{tabular}{ccccc}

    Parameter      & {Case A}  &{Case B}   &{Case C}  & {Search range}   \\ \hline
    $t_c/\rm yr$       & 3.5000   & 3.5000   & 3.5000  &[3.3, 3.8]       \\
    $M_c/M_{\odot}$ & 70.3032  & 54.3572  & 28.1923 &[20, 80]           \\
    $\delta_\mu$   & 0.1585   & 0.0400   & 0.2485  &[0, 0.9]           \\
    $\theta/\rm rad$   & 0.7854   & 0.7854   & 0.7854  &[0, $\pi$]         \\
    $\phi/\rm rad$     & 2.1216   & 2.1216   & 2.1216  & [0, $2\pi$]       \\
    $e_0(f_{\rm gw}=0.01\text{Hz})$ & 0.0100 & 0.0100 & 0.0100 &[0, 0.05]  \\
    $D_L/\rm Mpc$      & 920.9084 & 603.5043 & 123.9608& /                \\
    $\iota/\rm rad$    & 0.5329   & 0.5329   & 0.5329  & /                \\
    $\psi/\rm rad$     & 1.2501   & 1.2501   & 1.2501  & /                \\
    $\phi_c/\rm rad$   & 0.8419   & 0.8419   & 0.8419  & /                \\
    $\rm SNR$          & 11.3137  & 11.1723  & 14.1421 &/                 \\

    \end{tabular}
    \end{ruledtabular}
\end{table}

\begin{figure*}[htbt]
\includegraphics[width=1\textwidth, trim=20pt 35pt 20pt 10pt, clip]{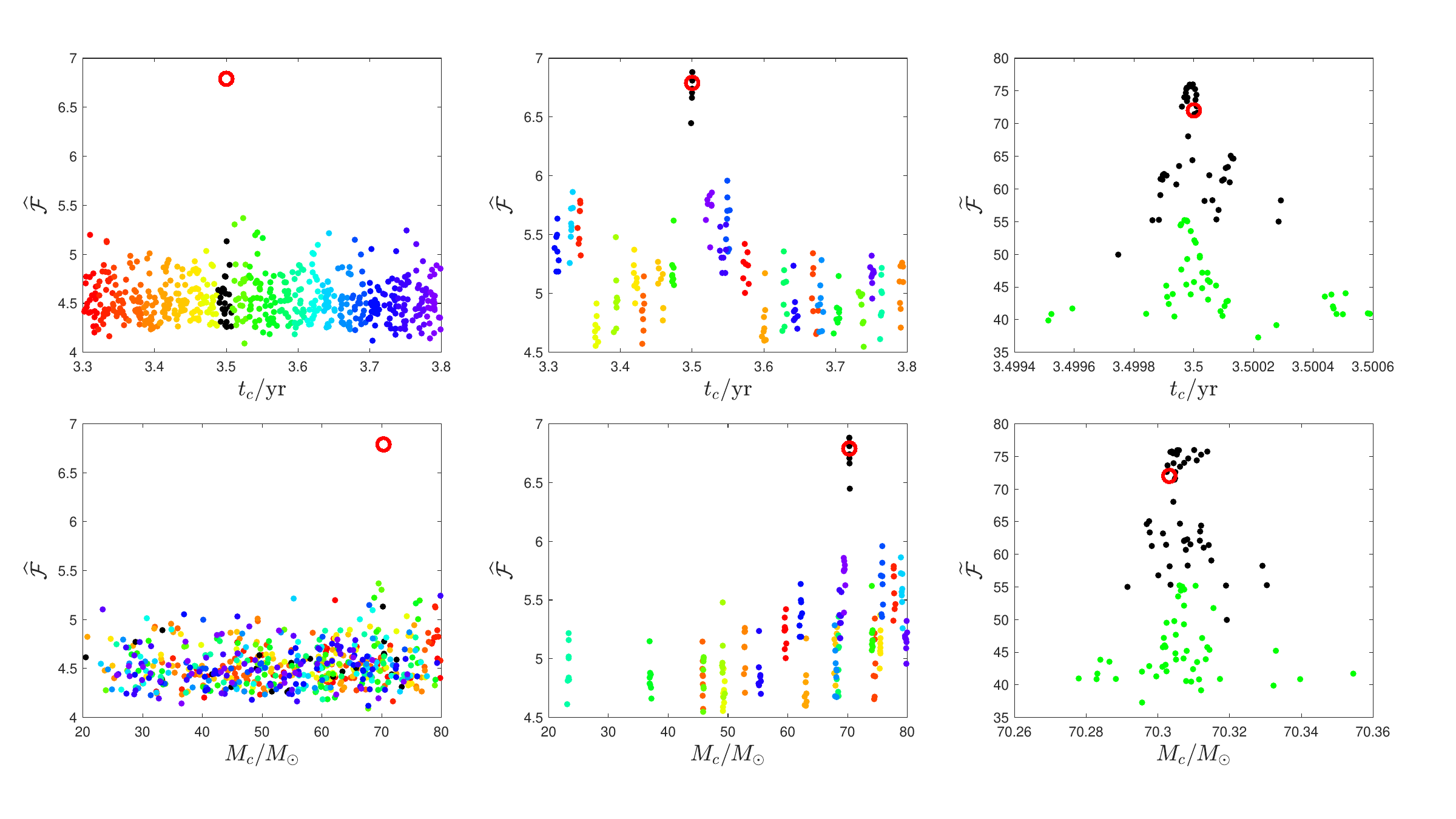}
\caption{\label{fig:10} Part of the PSO results about $t_c$ (top row) and $M_c$ (bottom row). In each panel, the red circle marks the true values of the parameters and their fitness, dots with different colors come from different search ranges of $t_c$. The black dots represent the PSO results from the range of $t_c$ that encloses its true value. From left to right are stage--1, stage--2, and stage--3.}
\end{figure*}

\begin{figure*}[htbt]
\includegraphics[width=1\textwidth, trim=20pt 35pt 20pt 10pt, clip]{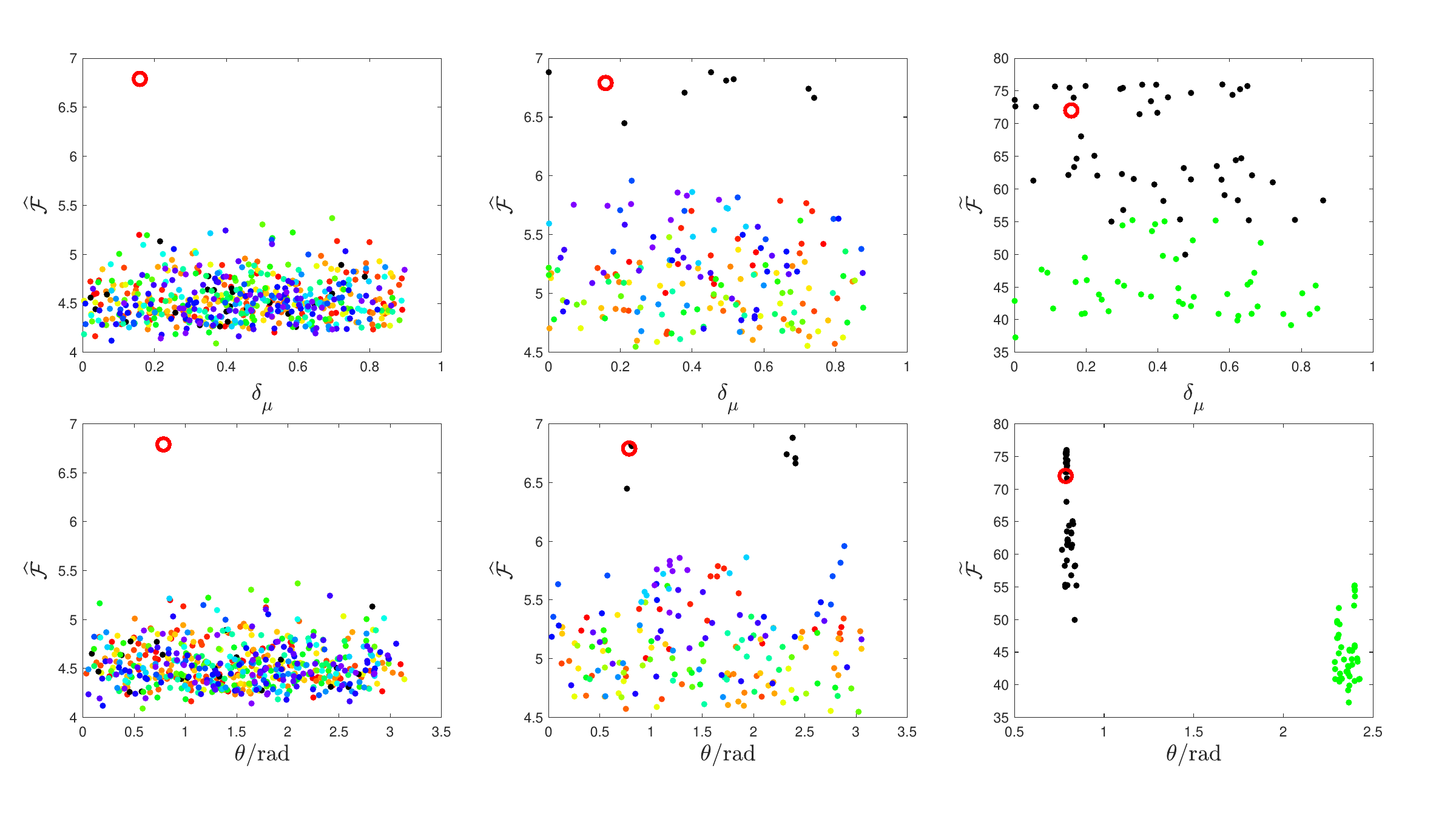}
\caption{\label{fig:11} As in Fig.~\ref{fig:10}, the PSO results about $\delta_\mu$ and $\theta$.}
\end{figure*}

\begin{figure*}[htbt]
\includegraphics[width=1\textwidth, trim=20pt 35pt 20pt 10pt, clip]{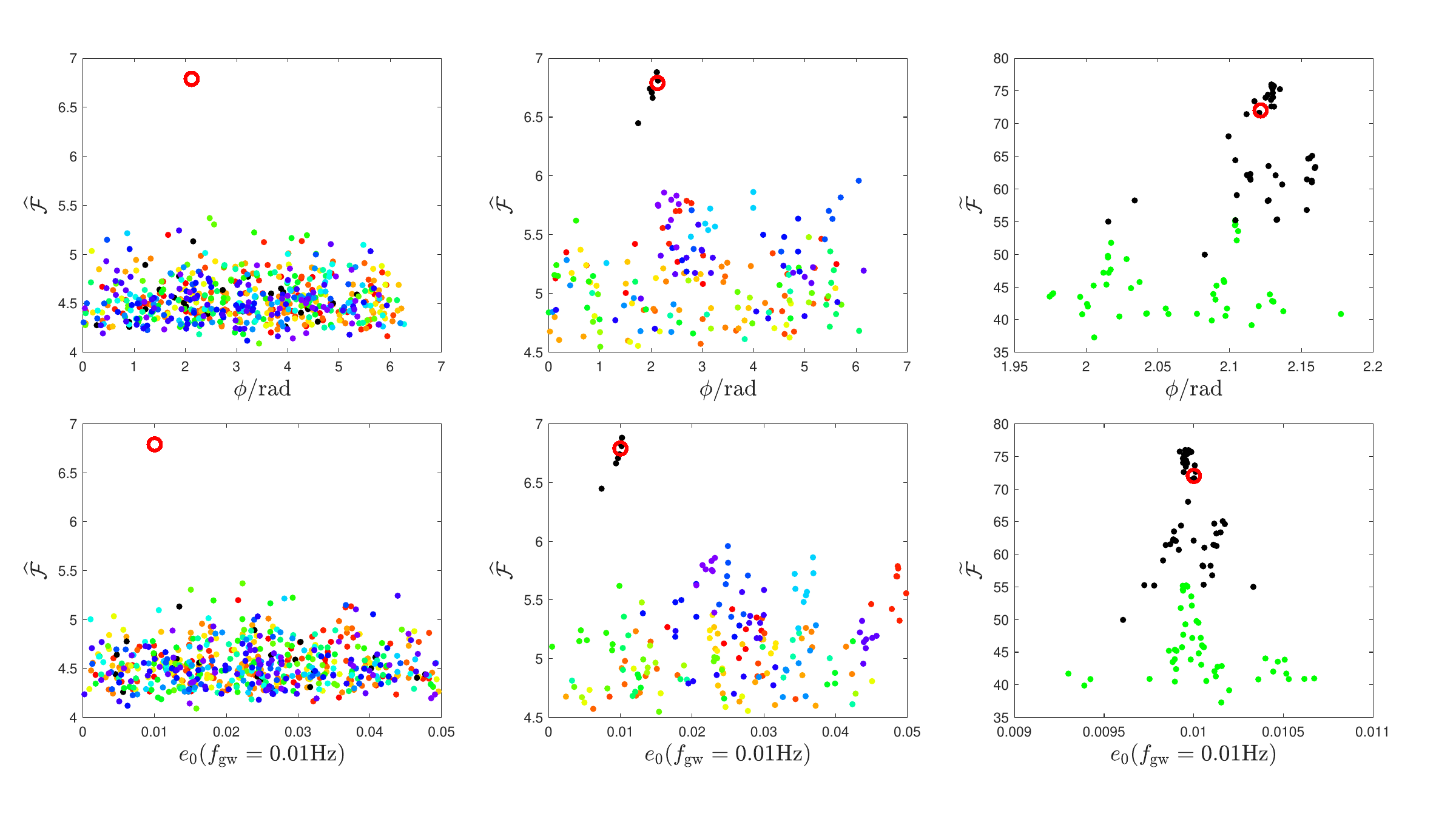}
\caption{\label{fig:12} As in Fig.~\ref{fig:10}, the PSO results about $\phi$ and $e_0$.}
\end{figure*}

The functioning of the hierarchical search is illustrated in Figures~\ref{fig:10} to~\ref{fig:12} pertaining to Case~A. Each figure shows the progression in the localization of the estimated parameter around the true value as one proceeds from stage--1 to stage--3. We see that there is a significant improvement for all parameters except $\delta_\mu$. The figures for stage--1 show clearly the effect of secondary maxima induced by noise. For the $t_c$ parameter, the secondary maxima are often larger than the semi-coherent $\mathcal{F}$-statistic values in the search range containing the true value of this parameter. 
It is worth noting that the latter may be just some values found by PSO, not necessarily local maxima at this range. They could be close to secondary maxima, but they could simply result from noise. 
As discussed earlier in Sec.~\ref{s3}, confining the stage--2 search to a subset of stage--1 search spaces based on the semi-coherent $\mathcal{F}$-statistic values could, therefore, lead to premature convergence. The results for stage--2 show the importance of restricting the search range in $M_c$ along with $t_c$. We see that, this leads to a significant improvement in the semi-coherent $\mathcal{F}$-statistic value and that it can often exceed the value at the true location. This is a positive indicator that PSO was able to successfully converge to the global maximum. Finally, we see how the search range is narrowed down by several orders of magnitude at the beginning of stage--3, thus allowing the fully-coherent $\mathcal{F}$-statistic to be computed successfully.

Tables~\ref{tab:CaseA} to~\ref{tab:CaseC} present the estimated parameters and their relative errors for the three cases. It can be seen that the method was able to estimate the intrinsic parameters with very high accuracy. For example, in Case A, the relative error in $M_c$ is less than $0.01\%$ and $t_c$ is determined with an accuracy of $\approx 100$~sec. The sky localization is accurate to within $0.2~{\rm deg}^2$, and the relative errors for eccentricity and chirp mass are less than $1\%$ in all cases. So far, we only provide point estimates for the unknown parameters by MLE, while reporting the uncertainties (error bars) of these estimates (interval estimates) is useful in actual data analysis. One way to obtain error bars in our method is to sample the likelihood function in the restricted parameter space used in Stage 3 using Markov Chain Monte Carlo, or to use a local Gaussian fit to the likelihood function~\cite{PhysRevD.85.123008}. This will be investigated in our future work.
\begin{table}[htbp]
    \centering
    \caption{Case A search results. The leftmost column lists the true parameters of the injected signal. The second column lists the 
    estimated values found from stage--3 and the third column presents the relative error between the true and estimated values of each parameter.}
    \label{tab:CaseA}
    \begin{ruledtabular}
    \begin{tabular}{cccc}
    
    Parameter    & {True value}  &{Estimated value}  &{Relative error}  \\ \hline
$t_c/\rm yr$       &3.500000 & 3.499997 & 0.0000899\% \\
$M_c/M_{\odot}$&70.3032 & 70.3101 & 0.00986\% \\
$\delta_\mu$   &0.1585 & 0.5797 & 265\% \\
$\theta/\rm rad$   &0.7854 & 0.7909 & 0.696\% \\
$\phi/\rm rad$     &2.1216 & 2.1291 & 0.354\% \\
$e_0(f_{\rm gw}=0.01\text{Hz})$&0.010000 & 0.009953 & 0.473\% \\
$D_L/\rm Mpc$      &920.9084 & 397.4851 & 56.8\% \\
$\iota/\rm rad$    &0.5329 & 1.2719 & 139\% \\
$\psi/\rm rad$     &1.2501 & -0.5077 & 141\% \\
$\phi_c/\rm rad$   &0.8419 & 0.1554 & 81.5\% \\

    \end{tabular}
    \end{ruledtabular}
\end{table}
\begin{table}[htbp]
    \centering
    \caption{Case B search results. The leftmost column lists the true parameters of the injected signal. The second column lists the 
    estimated values found from stage--3 and the third column presents the relative error between the true and estimated values of each parameter. }
    \label{tab:CaseB}
    \begin{ruledtabular}
    \begin{tabular}{cccc}
    
    Parameter    & {True value}  &{Estimated value}  &{Relative error}  \\ \hline
$t_c/\rm yr$       &3.5000000 & 3.5000142 & 0.000405\% \\
$M_c/M_{\odot}$&54.3572 & 54.3617 & 0.00838\% \\
$\delta_\mu$   &0.0400 & 0.5721 & 1330\% \\
$\theta/\rm rad$   &0.7854 & 0.7898 & 0.558\% \\
$\phi/\rm rad$     &2.1216 & 2.1192 & 0.112\% \\
$e_0(f_{\rm gw}=0.01\text{Hz})$&0.01000 & 0.009970 & 0.300\% \\
$D_L/\rm Mpc$      &603.5043 & 513.3089 & 14.9\% \\
$\iota/\rm rad$    &0.5329 & 0.7259 & 36.2\% \\
$\psi/\rm rad$     &1.2501 & 0.4386 & 64.9\% \\
$\phi_c/\rm rad$   &0.8419 & 0.4058 & 51.8\% \\

    \end{tabular}
    \end{ruledtabular}
\end{table}
\begin{table}[htbp]
    \centering
    \caption{Case C search results. The leftmost column lists the true parameters of the injected signal. The second column lists the 
    estimated values found from stage--3 and the third column presents the relative error between the true and estimated values of each parameter.}
    \label{tab:CaseC}
    \begin{ruledtabular}
    \begin{tabular}{cccc}
    
    Parameter    & {True value}  &{Estimated value}  &{Relative error}  \\ \hline
$t_c/\rm yr$       &3.500000 & 3.5000039 & 0.0001126\% \\
$M_c/M_{\odot}$&28.1923 & 28.1932 & 0.00315\% \\
$\delta_\mu$   &0.07692 & 0.41195 & 4355\% \\
$\theta/\rm rad$   &0.7854 & 0.7864 & 0.129\% \\
$\phi/\rm rad$     &2.1216 & 2.1228 & 0.0575\% \\
$e_0(f_{\rm gw}=0.01\text{Hz})$&0.0100 & 0.009961 & 0.388\% \\
$D_L/\rm Mpc$      &123.9608 & 74.1714 & 40.2\% \\
$\iota/\rm rad$    &0.5329 & 1.1470 & 115\% \\
$\psi/\rm rad$     &1.2501 & -0.3975 & 132\% \\
$\phi_c/\rm rad$   &0.8419 & 0.7437 & 11.7\% \\

    \end{tabular}
    \end{ruledtabular}
\end{table}

Case B and Case C differ from Case A primarily in the injected values of the chirp mass. This design allows us to validate our method at different points within its search range. The parameters $t_c$, $\theta$, $\phi$, and $e_{0}$, as well as the extrinsic parameters (except $D_L$), remain the same as in Case A. The parameter $D_L$ is adjusted to achieve the desired SNRs of the injected signals. As shown in Tables~\ref{tab:CaseB} to~\ref{tab:CaseC}, the relative errors in the intrinsic parameters are consistent with those in Case A, despite the noise realizations being independent draws based on the LISA instrumental noise PSD.

Moreover, the relative errors in $D_L$ exceed $10\%$ across all three cases with relatively low SNRs, posing a challenge for accurate distance measurement of these sources using GW alone. However, this may be mitigated by the high estimation accuracy achieved in $\theta$ and $\phi$, making it possible for electromagnetic follow-up to pinpoint their host galaxies. Additionally, predicting the time of coalescence of an SmBBH well before its final merger in high frequency band can provide valuable guidance to ground-based GW detectors. This allows the detectors to be adjusted to their optimal states for this source~\cite{tso2019optimizing}, thereby maximizing the scientific return by combining observations from both high- and low-frequency detectors.

\section{Conclusion}
\label{s6}
We have presented a hierarchical search method for SmBBH signals in noisy LISA data that shows promising performance for an SNR range of $\approx 11$ to $\approx 14$ for $2.5$~yr of a single TDI combination. The method uses a hierarchy consisting of partitioned parameter search ranges and semi-coherent $\mathcal{F}$-statistic with different numbers of segments. The search for the global maximum of the semi-coherent $\mathcal{F}$-statistic in each search range is performed using PSO. By employing frequency-domain responses and GPU parallel computing to accelerate the code, we were able to efficiently search for single GW signals within a reasonable timeframe of $2.5$~days. We validated the reliability of this pipeline using simulated data containing noise based on instrumental design power spectral density of LISA. Our results demonstrate that it is possible to address the data analysis challenges posed by SmBBH signals and produce highly accurate estimates of intrinsic parameters. In particular, the time of coalescence can be localized within $\approx 100$~sec and the sky location can be estimated within $\approx 0.2$~${\rm deg}^2$, which is more than adequate to enable multi-wavelength astronomy for SmBBH sources, and it also opens up the possibility of multi-messenger astronomy if there is an EM counterpart associated with the GW events.

LISA can also measure the orbital eccentricity during inspirals of the SmBBH with a relative error less than 1$\%$, providing the ability to distinguish between their origin from binary star evolution or the dynamical interactions in dense star clusters. However, our waveform model does not include higher-order modes that would be present in a system with high orbital eccentricity. In addition, our waveform model does not include the spins of the individual black holes, which may be important in some cases. We will address these limitations, including the use of multiple TDI combinations, in future studies.

Finally, the signals we simulated all merged shortly outside the observation time window, whereas in reality, some sources may merge within the window, while others may merge long after the window. This implies that the search range for the merger time should be expanded, and the problem of multiple sources overlapping should also be addressed.

\begin{acknowledgments}
We thank Xin-Chun Hu and Yi-Qian Qian for their helpful discussions. 
Y.W. gratefully acknowledges support from the National Key Research and Development Program of China (No. 2023YFC2206702 and No. 2022YFC2205201), the National Natural Science Foundation of China (NSFC) under Grant No. 11973024, Major Science and Technology Program of Xinjiang Uygur Autonomous Region (No. 2022A03013-4), and Guangdong Major Project of Basic and Applied Basic Research (Grant No. 2019B030302001). 
S.D.M. is supported by U.S. National Science Foundation (NSF) Grant No. PHY-2207935. 
We acknowledge the High Performance Computing Platform at Huazhong University of Science and Technology for providing computational resources. 
The authors thank the anonymous referee for helpful comments that have significantly improved our manuscript.

\end{acknowledgments}

\appendix

\section{Gravitational Waveforms}
\label{a1}
In this paper, we employ the restricted post-Newtonian waveforms in the frequency domain, where ``restricted" means that we neglect higher-order amplitude modulation and focus only on the post-Newtonian effects in the phase:
\begin{equation}
h(f) = Af^{-7/6} e^{-i\Psi(f)}
\end{equation}
where (for $f > 0)$:
\begin{equation}
A=-\sqrt{5/24} (M_c^{5/6})/(\pi^{2/3} D_L)  \,,
\end{equation}
\begin{equation}
\begin{aligned}
\Psi(f) &= 2\pi ft_c - \phi_c - \frac{\pi}{4} + \frac{3}{128} (M_c \pi f)^{-5/3} \sum_{k=0}^6 \alpha_k x^{k/2} \\
&- \phi_p(t(f)) - \phi_D(t(f)) + \phi_e(f) \,.
\end{aligned}
\end{equation}
The details of the parameters, including the coefficients $\alpha_k$, can be found in \cite{2019PhRvD..99l3002F,PhysRevD.57.7089,PhysRevD.52.2089,PhysRevD.80.084043}. In addition, for simplicity, we have omitted certain parameters, such as the spin of the individual black holes, from the waveforms in the current paper, resulting in a total of ten parameters represented by $\Theta$. 

\section{Relationships between extrinsic parameters and intrinsic parameters}
\label{a2}

In Sec.~\ref{s2.1}, the fully-coherent $\mathcal{F}$-statistic is obtained by maximizing the log-likelihood ratio function over $\Theta_E = \{a_i\}$, which are time-independent reparameterizations of $\{D_L, \psi, \iota, \phi_c\}$ as follows~\cite{2007PhRvD..75d3008C}: 
\begin{equation}
\begin{aligned}
a_1 = \frac{1}{D_L} \left( \cos 2\psi \frac{1 + \cos^2 \iota}{2} \cos \phi_c - \sin 2\psi \cos \iota \sin \phi_c \right) \,, \\ 
a_2 = -\frac{1}{D_L} \left( \sin 2\psi \frac{1 + \cos^2 \iota}{2} \cos \phi_c + \cos 2\psi \cos \iota \sin \phi_c \right) \,, \\
a_3 = -\frac{1}{D_L} \left( \cos 2\psi \frac{1 + \cos^2 \iota}{2} \sin \phi_c + \sin 2\psi \cos \iota \cos \phi_c \right) \,, \\
a_4 = \frac{1}{D_L} \left( \sin 2\psi \frac{1 + \cos^2 \iota}{2} \sin \phi_c - \cos 2\psi \cos \iota \cos \phi_c \right) \,.
\end{aligned}
\end{equation}

Once the maximum likelihood estimate of the intrinsic parameters ($\Theta_I$) is obtained, $a_i$ can be determined by Eq.~\ref{aandMN3}. Subsequently, the corresponding maximum likelihood estimate of $\{D_L, \psi, \iota, \phi_c\}$ can be obtained by the following relationship, 
\begin{equation}\label{eq:extrinPara}
\begin{aligned}
D_L &= \frac{2}{A}, \\
\iota &= \arccos\left(\frac{-A_{\times}}{A_{+} + \sqrt{A_{+}^2 - A_{\times}^2}}\right) \,, \\
\psi &= \frac{1}{2}\arctan\left(\frac{A_{+}a_4 - A_{\times}a_1}{-(A_{\times}a_2 + A_{+}a_3)}\right) \,, \\
\phi_c &= \arctan\left(\frac{c(A_{+}a_4 - A_{\times}a_1)}{-c(A_{\times}a_3 + A_{+}a_2)}\right) \,,
\end{aligned}
\end{equation}
where:
\begin{equation}
\begin{aligned}
A_{+} &= \sqrt{(a_1 + a_4)^2 + (a_2 - a_3)^2} + \sqrt{(a_1 - a_4)^2 + (a_2 + a_3)^2} \,, \\
A_{\times} &= \sqrt{(a_1 + a_4)^2 + (a_2 - a_3)^2} - \sqrt{(a_1 - a_4)^2 + (a_2 + a_3)^2} \,, \\
A &= A_{+} + \sqrt{A_{+}^2 - A_{\times}^2}, \\
c &= \text{sgn}(\sin(2\psi)) \,.
\end{aligned}
\end{equation}
%


\begin{thebibliography}{89}%
\makeatletter
\providecommand \@ifxundefined [1]{%
 \@ifx{#1\undefined}
}%
\providecommand \@ifnum [1]{%
 \ifnum #1\expandafter \@firstoftwo
 \else \expandafter \@secondoftwo
 \fi
}%
\providecommand \@ifx [1]{%
 \ifx #1\expandafter \@firstoftwo
 \else \expandafter \@secondoftwo
 \fi
}%
\providecommand \natexlab [1]{#1}%
\providecommand \enquote  [1]{``#1''}%
\providecommand \bibnamefont  [1]{#1}%
\providecommand \bibfnamefont [1]{#1}%
\providecommand \citenamefont [1]{#1}%
\providecommand \href@noop [0]{\@secondoftwo}%
\providecommand \href [0]{\begingroup \@sanitize@url \@href}%
\providecommand \@href[1]{\@@startlink{#1}\@@href}%
\providecommand \@@href[1]{\endgroup#1\@@endlink}%
\providecommand \@sanitize@url [0]{\catcode `\\12\catcode `\$12\catcode
  `\&12\catcode `\#12\catcode `\^12\catcode `\_12\catcode `\%12\relax}%
\providecommand \@@startlink[1]{}%
\providecommand \@@endlink[0]{}%
\providecommand \url  [0]{\begingroup\@sanitize@url \@url }%
\providecommand \@url [1]{\endgroup\@href {#1}{\urlprefix }}%
\providecommand \urlprefix  [0]{URL }%
\providecommand \Eprint [0]{\href }%
\providecommand \doibase [0]{https://doi.org/}%
\providecommand \selectlanguage [0]{\@gobble}%
\providecommand \bibinfo  [0]{\@secondoftwo}%
\providecommand \bibfield  [0]{\@secondoftwo}%
\providecommand \translation [1]{[#1]}%
\providecommand \BibitemOpen [0]{}%
\providecommand \bibitemStop [0]{}%
\providecommand \bibitemNoStop [0]{.\EOS\space}%
\providecommand \EOS [0]{\spacefactor3000\relax}%
\providecommand \BibitemShut  [1]{\csname bibitem#1\endcsname}%
\let\auto@bib@innerbib\@empty
\bibitem [{\citenamefont {Abbott}\ \emph
  {et~al.}(2016{\natexlab{a}})\citenamefont {Abbott}, \citenamefont {Abbott},
  \citenamefont {Abbott}, \citenamefont {Abernathy}, \citenamefont {Acernese}
  \emph {et~al.}}]{abbott2016observation}%
  \BibitemOpen
  \bibfield  {author} {\bibinfo {author} {\bibfnamefont {B.~P.}\ \bibnamefont
  {Abbott}}, \bibinfo {author} {\bibfnamefont {R.}~\bibnamefont {Abbott}},
  \bibinfo {author} {\bibfnamefont {T.}~\bibnamefont {Abbott}}, \bibinfo
  {author} {\bibfnamefont {M.}~\bibnamefont {Abernathy}}, \bibinfo {author}
  {\bibfnamefont {F.}~\bibnamefont {Acernese}}, \emph {et~al.},\ }\bibfield
  {title} {\bibinfo {title} {Observation of gravitational waves from a binary
  black hole merger},\ }\href {https://doi.org/10.1103/PhysRevLett.116.061102}
  {\bibfield  {journal} {\bibinfo  {journal} {Physical review letters}\
  }\textbf {\bibinfo {volume} {116}},\ \bibinfo {pages} {061102} (\bibinfo
  {year} {2016}{\natexlab{a}})}\BibitemShut {NoStop}%
\bibitem [{\citenamefont {Aasi}\ \emph {et~al.}(2015)\citenamefont {Aasi},
  \citenamefont {Abbott}, \citenamefont {Abbott}, \citenamefont {Abbott},
  \citenamefont {Abernathy} \emph {et~al.}}]{aasi2015advanced}%
  \BibitemOpen
  \bibfield  {author} {\bibinfo {author} {\bibfnamefont {J.}~\bibnamefont
  {Aasi}}, \bibinfo {author} {\bibfnamefont {B.}~\bibnamefont {Abbott}},
  \bibinfo {author} {\bibfnamefont {R.}~\bibnamefont {Abbott}}, \bibinfo
  {author} {\bibfnamefont {T.}~\bibnamefont {Abbott}}, \bibinfo {author}
  {\bibfnamefont {M.}~\bibnamefont {Abernathy}}, \emph {et~al.},\ }\bibfield
  {title} {\bibinfo {title} {Advanced ligo},\ }\href
  {https://doi.org/10.1088/0264-9381/32/7/074001} {\bibfield  {journal}
  {\bibinfo  {journal} {Classical and quantum gravity}\ }\textbf {\bibinfo
  {volume} {32}},\ \bibinfo {pages} {074001} (\bibinfo {year}
  {2015})}\BibitemShut {NoStop}%
\bibitem [{\citenamefont {{Acernese}}\ \emph {et~al.}(2015)\citenamefont
  {{Acernese}}, \citenamefont {{Agathos}}, \citenamefont {{Agatsuma}},
  \citenamefont {{Aisa}}, \citenamefont {{Allemandou}}, \citenamefont
  {{Allocca}}, \citenamefont {{Amarni}}, \citenamefont {{Astone}},
  \citenamefont {{Balestri}}, \citenamefont {{Ballardin}} \emph
  {et~al.}}]{2015CQGra..32b4001A}%
  \BibitemOpen
  \bibfield  {author} {\bibinfo {author} {\bibfnamefont {F.}~\bibnamefont
  {{Acernese}}}, \bibinfo {author} {\bibfnamefont {M.}~\bibnamefont
  {{Agathos}}}, \bibinfo {author} {\bibfnamefont {K.}~\bibnamefont
  {{Agatsuma}}}, \bibinfo {author} {\bibfnamefont {D.}~\bibnamefont {{Aisa}}},
  \bibinfo {author} {\bibfnamefont {N.}~\bibnamefont {{Allemandou}}}, \bibinfo
  {author} {\bibfnamefont {A.}~\bibnamefont {{Allocca}}}, \bibinfo {author}
  {\bibfnamefont {J.}~\bibnamefont {{Amarni}}}, \bibinfo {author}
  {\bibfnamefont {P.}~\bibnamefont {{Astone}}}, \bibinfo {author}
  {\bibfnamefont {G.}~\bibnamefont {{Balestri}}}, \bibinfo {author}
  {\bibfnamefont {G.}~\bibnamefont {{Ballardin}}}, \emph {et~al.},\ }\bibfield
  {title} {\bibinfo {title} {{Advanced Virgo: a second-generation
  interferometric gravitational wave detector}},\ }\href
  {https://doi.org/10.1088/0264-9381/32/2/024001} {\bibfield  {journal}
  {\bibinfo  {journal} {Classical and Quantum Gravity}\ }\textbf {\bibinfo
  {volume} {32}},\ \bibinfo {eid} {024001} (\bibinfo {year} {2015})},\ \Eprint
  {https://arxiv.org/abs/1408.3978} {arXiv:1408.3978 [gr-qc]} \BibitemShut
  {NoStop}%
\bibitem [{\citenamefont {{Somiya}}(2012)}]{2012CQGra..29l4007S}%
  \BibitemOpen
  \bibfield  {author} {\bibinfo {author} {\bibfnamefont {K.}~\bibnamefont
  {{Somiya}}},\ }\bibfield  {title} {\bibinfo {title} {{Detector configuration
  of KAGRA-the Japanese cryogenic gravitational-wave detector}},\ }\href
  {https://doi.org/10.1088/0264-9381/29/12/124007} {\bibfield  {journal}
  {\bibinfo  {journal} {Classical and Quantum Gravity}\ }\textbf {\bibinfo
  {volume} {29}},\ \bibinfo {eid} {124007} (\bibinfo {year} {2012})},\ \Eprint
  {https://arxiv.org/abs/1111.7185} {arXiv:1111.7185 [gr-qc]} \BibitemShut
  {NoStop}%
\bibitem [{\citenamefont {Abbott}\ \emph
  {et~al.}(2016{\natexlab{b}})\citenamefont {Abbott}, \citenamefont {Abbott},
  \citenamefont {Abbott}, \citenamefont {Abernathy}, \citenamefont {Acernese}
  \emph {et~al.}}]{abbott2016binary}%
  \BibitemOpen
  \bibfield  {author} {\bibinfo {author} {\bibfnamefont {B.~P.}\ \bibnamefont
  {Abbott}}, \bibinfo {author} {\bibfnamefont {R.}~\bibnamefont {Abbott}},
  \bibinfo {author} {\bibfnamefont {T.}~\bibnamefont {Abbott}}, \bibinfo
  {author} {\bibfnamefont {M.}~\bibnamefont {Abernathy}}, \bibinfo {author}
  {\bibfnamefont {F.}~\bibnamefont {Acernese}}, \emph {et~al.},\ }\bibfield
  {title} {\bibinfo {title} {Binary black hole mergers in the first advanced
  ligo observing run},\ }\href {https://doi.org/10.1103/PhysRevX.6.041015}
  {\bibfield  {journal} {\bibinfo  {journal} {Physical Review X}\ }\textbf
  {\bibinfo {volume} {6}},\ \bibinfo {pages} {041015} (\bibinfo {year}
  {2016}{\natexlab{b}})}\BibitemShut {NoStop}%
\bibitem [{\citenamefont {Abbott}\ \emph {et~al.}(2019)\citenamefont {Abbott},
  \citenamefont {Abbott}, \citenamefont {Abbott}, \citenamefont {Abraham},
  \citenamefont {Acernese} \emph {et~al.}}]{abbott2019gwtc}%
  \BibitemOpen
  \bibfield  {author} {\bibinfo {author} {\bibfnamefont {B.}~\bibnamefont
  {Abbott}}, \bibinfo {author} {\bibfnamefont {R.}~\bibnamefont {Abbott}},
  \bibinfo {author} {\bibfnamefont {T.}~\bibnamefont {Abbott}}, \bibinfo
  {author} {\bibfnamefont {S.}~\bibnamefont {Abraham}}, \bibinfo {author}
  {\bibfnamefont {F.}~\bibnamefont {Acernese}}, \emph {et~al.},\ }\bibfield
  {title} {\bibinfo {title} {Gwtc-1: a gravitational-wave transient catalog of
  compact binary mergers observed by ligo and virgo during the first and second
  observing runs},\ }\href {https://doi.org/10.1103/PhysRevX.9.031040}
  {\bibfield  {journal} {\bibinfo  {journal} {Physical Review X}\ }\textbf
  {\bibinfo {volume} {9}},\ \bibinfo {pages} {031040} (\bibinfo {year}
  {2019})}\BibitemShut {NoStop}%
\bibitem [{\citenamefont {Abbott}\ \emph {et~al.}(2021)\citenamefont {Abbott},
  \citenamefont {Abbott}, \citenamefont {Abraham}, \citenamefont {Acernese},
  \citenamefont {Ackley} \emph {et~al.}}]{abbott2021gwtc}%
  \BibitemOpen
  \bibfield  {author} {\bibinfo {author} {\bibfnamefont {R.}~\bibnamefont
  {Abbott}}, \bibinfo {author} {\bibfnamefont {T.}~\bibnamefont {Abbott}},
  \bibinfo {author} {\bibfnamefont {S.}~\bibnamefont {Abraham}}, \bibinfo
  {author} {\bibfnamefont {F.}~\bibnamefont {Acernese}}, \bibinfo {author}
  {\bibfnamefont {K.}~\bibnamefont {Ackley}}, \emph {et~al.},\ }\bibfield
  {title} {\bibinfo {title} {Gwtc-2: compact binary coalescences observed by
  ligo and virgo during the first half of the third observing run},\ }\href
  {https://doi.org/10.1103/PhysRevX.11.021053} {\bibfield  {journal} {\bibinfo
  {journal} {Physical Review X}\ }\textbf {\bibinfo {volume} {11}},\ \bibinfo
  {pages} {021053} (\bibinfo {year} {2021})}\BibitemShut {NoStop}%
\bibitem [{\citenamefont {Abbott}\ \emph {et~al.}(2023)\citenamefont {Abbott},
  \citenamefont {Abbott}, \citenamefont {Acernese}, \citenamefont {Ackley},
  \citenamefont {Adams} \emph {et~al.}}]{abbott2021gwtc-3}%
  \BibitemOpen
  \bibfield  {author} {\bibinfo {author} {\bibfnamefont {R.}~\bibnamefont
  {Abbott}}, \bibinfo {author} {\bibfnamefont {T.}~\bibnamefont {Abbott}},
  \bibinfo {author} {\bibfnamefont {F.}~\bibnamefont {Acernese}}, \bibinfo
  {author} {\bibfnamefont {K.}~\bibnamefont {Ackley}}, \bibinfo {author}
  {\bibfnamefont {C.}~\bibnamefont {Adams}}, \emph {et~al.},\ }\bibfield
  {title} {\bibinfo {title} {Gwtc-3: compact binary coalescences observed by
  ligo and virgo during the second part of the third observing run},\ }\href
  {https://doi.org/10.1103/PhysRevX.13.041039} {\bibfield  {journal} {\bibinfo
  {journal} {Physical Review X}\ }\textbf {\bibinfo {volume} {13}},\ \bibinfo
  {eid} {041039} (\bibinfo {year} {2023})},\ \Eprint
  {https://arxiv.org/abs/2111.03606} {arXiv:2111.03606 [gr-qc]} \BibitemShut
  {NoStop}%
\bibitem [{\citenamefont {{Punturo}}\ \emph {et~al.}(2010)\citenamefont
  {{Punturo}}, \citenamefont {{Abernathy}},\ and\ \citenamefont
  {et~al.}}]{ET2010}%
  \BibitemOpen
  \bibfield  {author} {\bibinfo {author} {\bibfnamefont {M.}~\bibnamefont
  {{Punturo}}}, \bibinfo {author} {\bibfnamefont {M.}~\bibnamefont
  {{Abernathy}}},\ and\ \bibinfo {author} {\bibnamefont {et~al.}},\ }\bibfield
  {title} {\bibinfo {title} {{The Einstein Telescope: a third-generation
  gravitational wave observatory}},\ }\href
  {https://doi.org/10.1088/0264-9381/27/19/194002} {\bibfield  {journal}
  {\bibinfo  {journal} {Classical and Quantum Gravity}\ }\textbf {\bibinfo
  {volume} {27}},\ \bibinfo {eid} {194002} (\bibinfo {year}
  {2010})}\BibitemShut {NoStop}%
\bibitem [{\citenamefont {{Srivastava}}\ \emph {et~al.}(2022)\citenamefont
  {{Srivastava}}, \citenamefont {{Davis}}, \citenamefont {{Kuns}},
  \citenamefont {{Landry}}, \citenamefont {{Ballmer}}, \citenamefont {{Evans}},
  \citenamefont {{Hall}}, \citenamefont {{Read}},\ and\ \citenamefont
  {{Sathyaprakash}}}]{CosmicExplorer2022}%
  \BibitemOpen
  \bibfield  {author} {\bibinfo {author} {\bibfnamefont {V.}~\bibnamefont
  {{Srivastava}}}, \bibinfo {author} {\bibfnamefont {D.}~\bibnamefont
  {{Davis}}}, \bibinfo {author} {\bibfnamefont {K.}~\bibnamefont {{Kuns}}},
  \bibinfo {author} {\bibfnamefont {P.}~\bibnamefont {{Landry}}}, \bibinfo
  {author} {\bibfnamefont {S.}~\bibnamefont {{Ballmer}}}, \bibinfo {author}
  {\bibfnamefont {M.}~\bibnamefont {{Evans}}}, \bibinfo {author} {\bibfnamefont
  {E.~D.}\ \bibnamefont {{Hall}}}, \bibinfo {author} {\bibfnamefont
  {J.}~\bibnamefont {{Read}}},\ and\ \bibinfo {author} {\bibfnamefont {B.~S.}\
  \bibnamefont {{Sathyaprakash}}},\ }\bibfield  {title} {\bibinfo {title}
  {{Science-driven Tunable Design of Cosmic Explorer Detectors}},\ }\href
  {https://doi.org/10.3847/1538-4357/ac5f04} {\bibfield  {journal} {\bibinfo
  {journal} {\apj}\ }\textbf {\bibinfo {volume} {931}},\ \bibinfo {eid} {22}
  (\bibinfo {year} {2022})},\ \Eprint {https://arxiv.org/abs/2201.10668}
  {arXiv:2201.10668 [gr-qc]} \BibitemShut {NoStop}%
\bibitem [{\citenamefont {{Hild}}\ \emph {et~al.}(2011)\citenamefont {{Hild}},
  \citenamefont {{Abernathy}}, \citenamefont {{Acernese}}, \citenamefont
  {{Amaro-Seoane}}, \citenamefont {{Andersson}},\ and\ \citenamefont
  {others.}}]{2011CQGra..28i4013H}%
  \BibitemOpen
  \bibfield  {author} {\bibinfo {author} {\bibfnamefont {S.}~\bibnamefont
  {{Hild}}}, \bibinfo {author} {\bibfnamefont {M.}~\bibnamefont {{Abernathy}}},
  \bibinfo {author} {\bibfnamefont {F.}~\bibnamefont {{Acernese}}}, \bibinfo
  {author} {\bibfnamefont {P.}~\bibnamefont {{Amaro-Seoane}}}, \bibinfo
  {author} {\bibfnamefont {N.}~\bibnamefont {{Andersson}}},\ and\ \bibinfo
  {author} {\bibnamefont {others.}},\ }\bibfield  {title} {\bibinfo {title}
  {{Sensitivity studies for third-generation gravitational wave
  observatories}},\ }\href {https://doi.org/10.1088/0264-9381/28/9/094013}
  {\bibfield  {journal} {\bibinfo  {journal} {Classical and Quantum Gravity}\
  }\textbf {\bibinfo {volume} {28}},\ \bibinfo {eid} {094013} (\bibinfo {year}
  {2011})},\ \Eprint {https://arxiv.org/abs/1012.0908} {arXiv:1012.0908
  [gr-qc]} \BibitemShut {NoStop}%
\bibitem [{\citenamefont {{Amaro-Seoane}}\ \emph {et~al.}(2017)\citenamefont
  {{Amaro-Seoane}}, \citenamefont {{Audley}}, \citenamefont {{Babak}},
  \citenamefont {{Baker}}, \citenamefont {{Barausse}}, \citenamefont
  {{Bender}}, \citenamefont {{Berti}}, \citenamefont {{Binetruy}},
  \citenamefont {{Born}}, \citenamefont {{Bortoluzzi}} \emph
  {et~al.}}]{2017arXiv170200786A}%
  \BibitemOpen
  \bibfield  {author} {\bibinfo {author} {\bibfnamefont {P.}~\bibnamefont
  {{Amaro-Seoane}}}, \bibinfo {author} {\bibfnamefont {H.}~\bibnamefont
  {{Audley}}}, \bibinfo {author} {\bibfnamefont {S.}~\bibnamefont {{Babak}}},
  \bibinfo {author} {\bibfnamefont {J.}~\bibnamefont {{Baker}}}, \bibinfo
  {author} {\bibfnamefont {E.}~\bibnamefont {{Barausse}}}, \bibinfo {author}
  {\bibfnamefont {P.}~\bibnamefont {{Bender}}}, \bibinfo {author}
  {\bibfnamefont {E.}~\bibnamefont {{Berti}}}, \bibinfo {author} {\bibfnamefont
  {P.}~\bibnamefont {{Binetruy}}}, \bibinfo {author} {\bibfnamefont
  {M.}~\bibnamefont {{Born}}}, \bibinfo {author} {\bibfnamefont
  {D.}~\bibnamefont {{Bortoluzzi}}}, \emph {et~al.},\ }\bibfield  {title}
  {\bibinfo {title} {{Laser Interferometer Space Antenna}},\ }\href
  {https://doi.org/10.48550/arXiv.1702.00786} {\bibfield  {journal} {\bibinfo
  {journal} {arXiv e-prints}\ ,\ \bibinfo {eid} {arXiv:1702.00786}} (\bibinfo
  {year} {2017})},\ \Eprint {https://arxiv.org/abs/1702.00786}
  {arXiv:1702.00786 [astro-ph.IM]} \BibitemShut {NoStop}%
\bibitem [{\citenamefont {{Luo}}\ \emph {et~al.}(2016)\citenamefont {{Luo}},
  \citenamefont {{Chen}}, \citenamefont {{Duan}}, \citenamefont {{Gong}},
  \citenamefont {{Hu}}, \citenamefont {{Ji}}, \citenamefont {{Liu}},
  \citenamefont {{Mei}}, \citenamefont {{Milyukov}}, \citenamefont {{Sazhin}}
  \emph {et~al.}}]{2016CQGra..33c5010L}%
  \BibitemOpen
  \bibfield  {author} {\bibinfo {author} {\bibfnamefont {J.}~\bibnamefont
  {{Luo}}}, \bibinfo {author} {\bibfnamefont {L.-S.}\ \bibnamefont {{Chen}}},
  \bibinfo {author} {\bibfnamefont {H.-Z.}\ \bibnamefont {{Duan}}}, \bibinfo
  {author} {\bibfnamefont {Y.-G.}\ \bibnamefont {{Gong}}}, \bibinfo {author}
  {\bibfnamefont {S.}~\bibnamefont {{Hu}}}, \bibinfo {author} {\bibfnamefont
  {J.}~\bibnamefont {{Ji}}}, \bibinfo {author} {\bibfnamefont {Q.}~\bibnamefont
  {{Liu}}}, \bibinfo {author} {\bibfnamefont {J.}~\bibnamefont {{Mei}}},
  \bibinfo {author} {\bibfnamefont {V.}~\bibnamefont {{Milyukov}}}, \bibinfo
  {author} {\bibfnamefont {M.}~\bibnamefont {{Sazhin}}}, \emph {et~al.},\
  }\bibfield  {title} {\bibinfo {title} {{TianQin: a space-borne gravitational
  wave detector}},\ }\href {https://doi.org/10.1088/0264-9381/33/3/035010}
  {\bibfield  {journal} {\bibinfo  {journal} {Classical and Quantum Gravity}\
  }\textbf {\bibinfo {volume} {33}},\ \bibinfo {eid} {035010} (\bibinfo {year}
  {2016})},\ \Eprint {https://arxiv.org/abs/1512.02076} {arXiv:1512.02076
  [astro-ph.IM]} \BibitemShut {NoStop}%
\bibitem [{\citenamefont {{Mei}}\ \emph {et~al.}(2021)\citenamefont {{Mei}},
  \citenamefont {{Bai}}, \citenamefont {{Bao}}, \citenamefont {{Barausse}},
  \citenamefont {{Cai}},\ and\ \citenamefont {others.}}]{2021PTEP.2021eA107M}%
  \BibitemOpen
  \bibfield  {author} {\bibinfo {author} {\bibfnamefont {J.}~\bibnamefont
  {{Mei}}}, \bibinfo {author} {\bibfnamefont {Y.-Z.}\ \bibnamefont {{Bai}}},
  \bibinfo {author} {\bibfnamefont {J.}~\bibnamefont {{Bao}}}, \bibinfo
  {author} {\bibfnamefont {E.}~\bibnamefont {{Barausse}}}, \bibinfo {author}
  {\bibfnamefont {L.}~\bibnamefont {{Cai}}},\ and\ \bibinfo {author}
  {\bibnamefont {others.}},\ }\bibfield  {title} {\bibinfo {title} {{The
  TianQin project: Current progress on science and technology}},\ }\href
  {https://doi.org/10.1093/ptep/ptaa114} {\bibfield  {journal} {\bibinfo
  {journal} {Progress of Theoretical and Experimental Physics}\ }\textbf
  {\bibinfo {volume} {2021}},\ \bibinfo {eid} {05A107} (\bibinfo {year}
  {2021})},\ \Eprint {https://arxiv.org/abs/2008.10332} {arXiv:2008.10332
  [gr-qc]} \BibitemShut {NoStop}%
\bibitem [{\citenamefont {Hu}\ and\ \citenamefont
  {Wu}(2017)}]{10.1093/nsr/nwx116}%
  \BibitemOpen
  \bibfield  {author} {\bibinfo {author} {\bibfnamefont {W.-R.}\ \bibnamefont
  {Hu}}\ and\ \bibinfo {author} {\bibfnamefont {Y.-L.}\ \bibnamefont {Wu}},\
  }\bibfield  {title} {\bibinfo {title} {{The Taiji Program in Space for
  gravitational wave physics and the nature of gravity}},\ }\href
  {https://doi.org/10.1093/nsr/nwx116} {\bibfield  {journal} {\bibinfo
  {journal} {National Science Review}\ }\textbf {\bibinfo {volume} {4}},\
  \bibinfo {pages} {685} (\bibinfo {year} {2017})},\ \Eprint
  {https://arxiv.org/abs/https://academic.oup.com/nsr/article-pdf/4/5/685/31566708/nwx116.pdf}
  {https://academic.oup.com/nsr/article-pdf/4/5/685/31566708/nwx116.pdf}
  \BibitemShut {NoStop}%
\bibitem [{\citenamefont {{Ruan}}\ \emph {et~al.}(2020)\citenamefont {{Ruan}},
  \citenamefont {{Guo}}, \citenamefont {{Cai}},\ and\ \citenamefont
  {{Zhang}}}]{2020IJMPA..3550075R}%
  \BibitemOpen
  \bibfield  {author} {\bibinfo {author} {\bibfnamefont {W.-H.}\ \bibnamefont
  {{Ruan}}}, \bibinfo {author} {\bibfnamefont {Z.-K.}\ \bibnamefont {{Guo}}},
  \bibinfo {author} {\bibfnamefont {R.-G.}\ \bibnamefont {{Cai}}},\ and\
  \bibinfo {author} {\bibfnamefont {Y.-Z.}\ \bibnamefont {{Zhang}}},\
  }\bibfield  {title} {\bibinfo {title} {{Taiji program: Gravitational-wave
  sources}},\ }\href {https://doi.org/10.1142/S0217751X2050075X} {\bibfield
  {journal} {\bibinfo  {journal} {International Journal of Modern Physics A}\
  }\textbf {\bibinfo {volume} {35}},\ \bibinfo {eid} {2050075} (\bibinfo {year}
  {2020})}\BibitemShut {NoStop}%
\bibitem [{\citenamefont {{Zhang}}\ \emph {et~al.}(2022)\citenamefont
  {{Zhang}}, \citenamefont {{Zhao}}, \citenamefont {{Mohanty}},\ and\
  \citenamefont {{Liu}}}]{2022PhRvD.106j2004Z}%
  \BibitemOpen
  \bibfield  {author} {\bibinfo {author} {\bibfnamefont {X.-H.}\ \bibnamefont
  {{Zhang}}}, \bibinfo {author} {\bibfnamefont {S.-D.}\ \bibnamefont {{Zhao}}},
  \bibinfo {author} {\bibfnamefont {S.~D.}\ \bibnamefont {{Mohanty}}},\ and\
  \bibinfo {author} {\bibfnamefont {Y.-X.}\ \bibnamefont {{Liu}}},\ }\bibfield
  {title} {\bibinfo {title} {{Resolving Galactic binaries using a network of
  space-borne gravitational wave detectors}},\ }\href
  {https://doi.org/10.1103/PhysRevD.106.102004} {\bibfield  {journal} {\bibinfo
   {journal} {\prd}\ }\textbf {\bibinfo {volume} {106}},\ \bibinfo {eid}
  {102004} (\bibinfo {year} {2022})},\ \Eprint
  {https://arxiv.org/abs/2206.12083} {arXiv:2206.12083 [gr-qc]} \BibitemShut
  {NoStop}%
\bibitem [{\citenamefont {{Gong}}\ \emph {et~al.}(2021)\citenamefont {{Gong}},
  \citenamefont {{Luo}},\ and\ \citenamefont {{Wang}}}]{2021NatAs...5..881G}%
  \BibitemOpen
  \bibfield  {author} {\bibinfo {author} {\bibfnamefont {Y.}~\bibnamefont
  {{Gong}}}, \bibinfo {author} {\bibfnamefont {J.}~\bibnamefont {{Luo}}},\ and\
  \bibinfo {author} {\bibfnamefont {B.}~\bibnamefont {{Wang}}},\ }\bibfield
  {title} {\bibinfo {title} {{Concepts and status of Chinese space
  gravitational wave detection projects}},\ }\href
  {https://doi.org/10.1038/s41550-021-01480-3} {\bibfield  {journal} {\bibinfo
  {journal} {Nature Astronomy}\ }\textbf {\bibinfo {volume} {5}},\ \bibinfo
  {pages} {881} (\bibinfo {year} {2021})},\ \Eprint
  {https://arxiv.org/abs/2109.07442} {arXiv:2109.07442 [astro-ph.IM]}
  \BibitemShut {NoStop}%
\bibitem [{\citenamefont {{Torres-Orjuela}}\ \emph {et~al.}(2023)\citenamefont
  {{Torres-Orjuela}}, \citenamefont {{Huang}}, \citenamefont {{Liang}},
  \citenamefont {{Liu}}, \citenamefont {{Wang}} \emph
  {et~al.}}]{2023arXiv230716628T}%
  \BibitemOpen
  \bibfield  {author} {\bibinfo {author} {\bibfnamefont {A.}~\bibnamefont
  {{Torres-Orjuela}}}, \bibinfo {author} {\bibfnamefont {S.-J.}\ \bibnamefont
  {{Huang}}}, \bibinfo {author} {\bibfnamefont {Z.-C.}\ \bibnamefont
  {{Liang}}}, \bibinfo {author} {\bibfnamefont {S.}~\bibnamefont {{Liu}}},
  \bibinfo {author} {\bibfnamefont {H.-T.}\ \bibnamefont {{Wang}}}, \emph
  {et~al.},\ }\bibfield  {title} {\bibinfo {title} {{Detection of astrophysical
  gravitational wave sources by TianQin and LISA}},\ }\href
  {https://doi.org/10.48550/arXiv.2307.16628} {\bibfield  {journal} {\bibinfo
  {journal} {arXiv e-prints}\ ,\ \bibinfo {eid} {arXiv:2307.16628}} (\bibinfo
  {year} {2023})},\ \Eprint {https://arxiv.org/abs/2307.16628}
  {arXiv:2307.16628 [gr-qc]} \BibitemShut {NoStop}%
\bibitem [{\citenamefont {{Jin}}\ \emph {et~al.}(2023)\citenamefont {{Jin}},
  \citenamefont {{Zhang}}, \citenamefont {{Song}}, \citenamefont {{Zhang}},\
  and\ \citenamefont {{Zhang}}}]{2023arXiv230519714J}%
  \BibitemOpen
  \bibfield  {author} {\bibinfo {author} {\bibfnamefont {S.-J.}\ \bibnamefont
  {{Jin}}}, \bibinfo {author} {\bibfnamefont {Y.-Z.}\ \bibnamefont {{Zhang}}},
  \bibinfo {author} {\bibfnamefont {J.-Y.}\ \bibnamefont {{Song}}}, \bibinfo
  {author} {\bibfnamefont {J.-F.}\ \bibnamefont {{Zhang}}},\ and\ \bibinfo
  {author} {\bibfnamefont {X.}~\bibnamefont {{Zhang}}},\ }\bibfield  {title}
  {\bibinfo {title} {{The Taiji-TianQin-LISA network: Precisely measuring the
  Hubble constant using both bright and dark sirens}},\ }\href
  {https://doi.org/10.48550/arXiv.2305.19714} {\bibfield  {journal} {\bibinfo
  {journal} {arXiv e-prints}\ ,\ \bibinfo {eid} {arXiv:2305.19714}} (\bibinfo
  {year} {2023})},\ \Eprint {https://arxiv.org/abs/2305.19714}
  {arXiv:2305.19714 [astro-ph.CO]} \BibitemShut {NoStop}%
\bibitem [{\citenamefont {{Amaro-Seoane}}\ \emph {et~al.}(2023)\citenamefont
  {{Amaro-Seoane}}, \citenamefont {{Andrews}}, \citenamefont {{Arca Sedda}},
  \citenamefont {{Askar}}, \citenamefont {{Baghi}} \emph
  {et~al.}}]{2023LRR....26....2A}%
  \BibitemOpen
  \bibfield  {author} {\bibinfo {author} {\bibfnamefont {P.}~\bibnamefont
  {{Amaro-Seoane}}}, \bibinfo {author} {\bibfnamefont {J.}~\bibnamefont
  {{Andrews}}}, \bibinfo {author} {\bibfnamefont {M.}~\bibnamefont {{Arca
  Sedda}}}, \bibinfo {author} {\bibfnamefont {A.}~\bibnamefont {{Askar}}},
  \bibinfo {author} {\bibfnamefont {Q.}~\bibnamefont {{Baghi}}}, \emph
  {et~al.},\ }\bibfield  {title} {\bibinfo {title} {{Astrophysics with the
  Laser Interferometer Space Antenna}},\ }\href
  {https://doi.org/10.1007/s41114-022-00041-y} {\bibfield  {journal} {\bibinfo
  {journal} {Living Reviews in Relativity}\ }\textbf {\bibinfo {volume} {26}},\
  \bibinfo {eid} {2} (\bibinfo {year} {2023})},\ \Eprint
  {https://arxiv.org/abs/2203.06016} {arXiv:2203.06016 [gr-qc]} \BibitemShut
  {NoStop}%
\bibitem [{\citenamefont {{Arun}}\ \emph {et~al.}(2022)\citenamefont {{Arun}},
  \citenamefont {{Belgacem}}, \citenamefont {{Benkel}}, \citenamefont
  {{Bernard}}, \citenamefont {{Berti}} \emph {et~al.}}]{2022LRR....25....4A}%
  \BibitemOpen
  \bibfield  {author} {\bibinfo {author} {\bibfnamefont {K.~G.}\ \bibnamefont
  {{Arun}}}, \bibinfo {author} {\bibfnamefont {E.}~\bibnamefont {{Belgacem}}},
  \bibinfo {author} {\bibfnamefont {R.}~\bibnamefont {{Benkel}}}, \bibinfo
  {author} {\bibfnamefont {L.}~\bibnamefont {{Bernard}}}, \bibinfo {author}
  {\bibfnamefont {E.}~\bibnamefont {{Berti}}}, \emph {et~al.},\ }\bibfield
  {title} {\bibinfo {title} {{New horizons for fundamental physics with
  LISA}},\ }\href {https://doi.org/10.1007/s41114-022-00036-9} {\bibfield
  {journal} {\bibinfo  {journal} {Living Reviews in Relativity}\ }\textbf
  {\bibinfo {volume} {25}},\ \bibinfo {eid} {4} (\bibinfo {year} {2022})},\
  \Eprint {https://arxiv.org/abs/2205.01597} {arXiv:2205.01597 [gr-qc]}
  \BibitemShut {NoStop}%
\bibitem [{\citenamefont {Sesana}(2016)}]{sesana2016prospects}%
  \BibitemOpen
  \bibfield  {author} {\bibinfo {author} {\bibfnamefont {A.}~\bibnamefont
  {Sesana}},\ }\bibfield  {title} {\bibinfo {title} {Prospects for multiband
  gravitational-wave astronomy after gw150914},\ }\href
  {https://doi.org/10.1103/PhysRevLett.116.231102} {\bibfield  {journal}
  {\bibinfo  {journal} {Physical Review Letters}\ }\textbf {\bibinfo {volume}
  {116}},\ \bibinfo {pages} {231102} (\bibinfo {year} {2016})}\BibitemShut
  {NoStop}%
\bibitem [{\citenamefont {Kyutoku}\ and\ \citenamefont
  {Seto}(2016)}]{kyutoku2016concise}%
  \BibitemOpen
  \bibfield  {author} {\bibinfo {author} {\bibfnamefont {K.}~\bibnamefont
  {Kyutoku}}\ and\ \bibinfo {author} {\bibfnamefont {N.}~\bibnamefont {Seto}},\
  }\bibfield  {title} {\bibinfo {title} {Concise estimate of the expected
  number of detections for stellar-mass binary black holes by elisa},\ }\href
  {https://doi.org/10.1093/mnras/stw1767} {\bibfield  {journal} {\bibinfo
  {journal} {Monthly Notices of the Royal Astronomical Society}\ }\textbf
  {\bibinfo {volume} {462}},\ \bibinfo {pages} {2177} (\bibinfo {year}
  {2016})}\BibitemShut {NoStop}%
\bibitem [{\citenamefont {Wong}\ \emph {et~al.}(2018)\citenamefont {Wong},
  \citenamefont {Kovetz}, \citenamefont {Cutler},\ and\ \citenamefont
  {Berti}}]{wong2018expanding}%
  \BibitemOpen
  \bibfield  {author} {\bibinfo {author} {\bibfnamefont {K.~W.}\ \bibnamefont
  {Wong}}, \bibinfo {author} {\bibfnamefont {E.~D.}\ \bibnamefont {Kovetz}},
  \bibinfo {author} {\bibfnamefont {C.}~\bibnamefont {Cutler}},\ and\ \bibinfo
  {author} {\bibfnamefont {E.}~\bibnamefont {Berti}},\ }\bibfield  {title}
  {\bibinfo {title} {Expanding the lisa horizon from the ground},\ }\href
  {https://doi.org/10.1103/PhysRevLett.121.251102} {\bibfield  {journal}
  {\bibinfo  {journal} {Physical review letters}\ }\textbf {\bibinfo {volume}
  {121}},\ \bibinfo {pages} {251102} (\bibinfo {year} {2018})}\BibitemShut
  {NoStop}%
\bibitem [{\citenamefont {Moore}\ \emph {et~al.}(2019)\citenamefont {Moore},
  \citenamefont {Gerosa},\ and\ \citenamefont {Klein}}]{moore2019stellar}%
  \BibitemOpen
  \bibfield  {author} {\bibinfo {author} {\bibfnamefont {C.~J.}\ \bibnamefont
  {Moore}}, \bibinfo {author} {\bibfnamefont {D.}~\bibnamefont {Gerosa}},\ and\
  \bibinfo {author} {\bibfnamefont {A.}~\bibnamefont {Klein}},\ }\bibfield
  {title} {\bibinfo {title} {Are stellar-mass black-hole binaries too quiet for
  lisa?},\ }\href {https://doi.org/10.1093/mnrasl/slz104} {\bibfield  {journal}
  {\bibinfo  {journal} {Monthly Notices of the Royal Astronomical Society:
  Letters}\ }\textbf {\bibinfo {volume} {488}},\ \bibinfo {pages} {L94}
  (\bibinfo {year} {2019})}\BibitemShut {NoStop}%
\bibitem [{\citenamefont {{Sato}}\ \emph {et~al.}(2017)\citenamefont {{Sato}},
  \citenamefont {{Kawamura}}, \citenamefont {{Ando}}, \citenamefont
  {{Nakamura}}, \citenamefont {{Tsubono}}, \citenamefont {{Araya}},
  \citenamefont {{Funaki}}, \citenamefont {{Ioka}}, \citenamefont {{Kanda}},
  \citenamefont {{Moriwaki}} \emph {et~al.}}]{2017JPhCS.840a2010S}%
  \BibitemOpen
  \bibfield  {author} {\bibinfo {author} {\bibfnamefont {S.}~\bibnamefont
  {{Sato}}}, \bibinfo {author} {\bibfnamefont {S.}~\bibnamefont {{Kawamura}}},
  \bibinfo {author} {\bibfnamefont {M.}~\bibnamefont {{Ando}}}, \bibinfo
  {author} {\bibfnamefont {T.}~\bibnamefont {{Nakamura}}}, \bibinfo {author}
  {\bibfnamefont {K.}~\bibnamefont {{Tsubono}}}, \bibinfo {author}
  {\bibfnamefont {A.}~\bibnamefont {{Araya}}}, \bibinfo {author} {\bibfnamefont
  {I.}~\bibnamefont {{Funaki}}}, \bibinfo {author} {\bibfnamefont
  {K.}~\bibnamefont {{Ioka}}}, \bibinfo {author} {\bibfnamefont
  {N.}~\bibnamefont {{Kanda}}}, \bibinfo {author} {\bibfnamefont
  {S.}~\bibnamefont {{Moriwaki}}}, \emph {et~al.},\ }\bibfield  {title}
  {\bibinfo {title} {{The status of DECIGO}},\ }in\ \href
  {https://doi.org/10.1088/1742-6596/840/1/012010} {\emph {\bibinfo {booktitle}
  {Journal of Physics Conference Series}}},\ \bibinfo {series} {Journal of
  Physics Conference Series}, Vol.\ \bibinfo {volume} {840}\ (\bibinfo {year}
  {2017})\ p.\ \bibinfo {pages} {012010}\BibitemShut {NoStop}%
\bibitem [{\citenamefont {{Kawamura}}\ \emph {et~al.}(2021)\citenamefont
  {{Kawamura}}, \citenamefont {{Ando}}, \citenamefont {{Seto}}, \citenamefont
  {{Sato}}, \citenamefont {{Musha}}, \citenamefont {{Kawano}}, \citenamefont
  {{Yokoyama}}, \citenamefont {{Tanaka}}, \citenamefont {{Ioka}}, \citenamefont
  {{Akutsu}} \emph {et~al.}}]{2021PTEP.2021eA105K}%
  \BibitemOpen
  \bibfield  {author} {\bibinfo {author} {\bibfnamefont {S.}~\bibnamefont
  {{Kawamura}}}, \bibinfo {author} {\bibfnamefont {M.}~\bibnamefont {{Ando}}},
  \bibinfo {author} {\bibfnamefont {N.}~\bibnamefont {{Seto}}}, \bibinfo
  {author} {\bibfnamefont {S.}~\bibnamefont {{Sato}}}, \bibinfo {author}
  {\bibfnamefont {M.}~\bibnamefont {{Musha}}}, \bibinfo {author} {\bibfnamefont
  {I.}~\bibnamefont {{Kawano}}}, \bibinfo {author} {\bibfnamefont
  {J.}~\bibnamefont {{Yokoyama}}}, \bibinfo {author} {\bibfnamefont
  {T.}~\bibnamefont {{Tanaka}}}, \bibinfo {author} {\bibfnamefont
  {K.}~\bibnamefont {{Ioka}}}, \bibinfo {author} {\bibfnamefont
  {T.}~\bibnamefont {{Akutsu}}}, \emph {et~al.},\ }\bibfield  {title} {\bibinfo
  {title} {{Current status of space gravitational wave antenna DECIGO and
  B-DECIGO}},\ }\href {https://doi.org/10.1093/ptep/ptab019} {\bibfield
  {journal} {\bibinfo  {journal} {Progress of Theoretical and Experimental
  Physics}\ }\textbf {\bibinfo {volume} {2021}},\ \bibinfo {eid} {05A105}
  (\bibinfo {year} {2021})},\ \Eprint {https://arxiv.org/abs/2006.13545}
  {arXiv:2006.13545 [gr-qc]} \BibitemShut {NoStop}%
\bibitem [{\citenamefont {Liu}\ \emph {et~al.}(2020)\citenamefont {Liu},
  \citenamefont {Hu}, \citenamefont {Zhang}, \citenamefont {Mei} \emph
  {et~al.}}]{liu2020science}%
  \BibitemOpen
  \bibfield  {author} {\bibinfo {author} {\bibfnamefont {S.}~\bibnamefont
  {Liu}}, \bibinfo {author} {\bibfnamefont {Y.-M.}\ \bibnamefont {Hu}},
  \bibinfo {author} {\bibfnamefont {J.-d.}\ \bibnamefont {Zhang}}, \bibinfo
  {author} {\bibfnamefont {J.}~\bibnamefont {Mei}}, \emph {et~al.},\ }\bibfield
   {title} {\bibinfo {title} {Science with the tianqin observatory: Preliminary
  results on stellar-mass binary black holes},\ }\href
  {https://doi.org/10.1103/PhysRevD.101.103027} {\bibfield  {journal} {\bibinfo
   {journal} {Physical Review D}\ }\textbf {\bibinfo {volume} {101}},\ \bibinfo
  {pages} {103027} (\bibinfo {year} {2020})}\BibitemShut {NoStop}%
\bibitem [{\citenamefont {Nishizawa}\ \emph {et~al.}(2016)\citenamefont
  {Nishizawa}, \citenamefont {Berti}, \citenamefont {Klein},\ and\
  \citenamefont {Sesana}}]{nishizawa2016elisa}%
  \BibitemOpen
  \bibfield  {author} {\bibinfo {author} {\bibfnamefont {A.}~\bibnamefont
  {Nishizawa}}, \bibinfo {author} {\bibfnamefont {E.}~\bibnamefont {Berti}},
  \bibinfo {author} {\bibfnamefont {A.}~\bibnamefont {Klein}},\ and\ \bibinfo
  {author} {\bibfnamefont {A.}~\bibnamefont {Sesana}},\ }\bibfield  {title}
  {\bibinfo {title} {elisa eccentricity measurements as tracers of binary black
  hole formation},\ }\href {https://doi.org/10.1103/PhysRevD.94.064020}
  {\bibfield  {journal} {\bibinfo  {journal} {Physical Review D}\ }\textbf
  {\bibinfo {volume} {94}},\ \bibinfo {pages} {064020} (\bibinfo {year}
  {2016})}\BibitemShut {NoStop}%
\bibitem [{\citenamefont {Breivik}\ \emph {et~al.}(2016)\citenamefont
  {Breivik}, \citenamefont {Rodriguez}, \citenamefont {Larson}, \citenamefont
  {Kalogera},\ and\ \citenamefont {Rasio}}]{breivik2016distinguishing}%
  \BibitemOpen
  \bibfield  {author} {\bibinfo {author} {\bibfnamefont {K.}~\bibnamefont
  {Breivik}}, \bibinfo {author} {\bibfnamefont {C.~L.}\ \bibnamefont
  {Rodriguez}}, \bibinfo {author} {\bibfnamefont {S.~L.}\ \bibnamefont
  {Larson}}, \bibinfo {author} {\bibfnamefont {V.}~\bibnamefont {Kalogera}},\
  and\ \bibinfo {author} {\bibfnamefont {F.~A.}\ \bibnamefont {Rasio}},\
  }\bibfield  {title} {\bibinfo {title} {Distinguishing between formation
  channels for binary black holes with lisa},\ }\href
  {https://doi.org/10.3847/2041-8205/830/1/L18} {\bibfield  {journal} {\bibinfo
   {journal} {The Astrophysical Journal Letters}\ }\textbf {\bibinfo {volume}
  {830}},\ \bibinfo {pages} {L18} (\bibinfo {year} {2016})}\BibitemShut
  {NoStop}%
\bibitem [{\citenamefont {Samsing}\ and\ \citenamefont
  {D’Orazio}(2018)}]{samsing2018black}%
  \BibitemOpen
  \bibfield  {author} {\bibinfo {author} {\bibfnamefont {J.}~\bibnamefont
  {Samsing}}\ and\ \bibinfo {author} {\bibfnamefont {D.~J.}\ \bibnamefont
  {D’Orazio}},\ }\bibfield  {title} {\bibinfo {title} {Black hole mergers
  from globular clusters observable by lisa i: eccentric sources originating
  from relativistic n-body dynamics},\ }\href
  {https://doi.org/10.1093/mnras/sty2334} {\bibfield  {journal} {\bibinfo
  {journal} {Monthly Notices of the Royal Astronomical Society}\ }\textbf
  {\bibinfo {volume} {481}},\ \bibinfo {pages} {5445} (\bibinfo {year}
  {2018})}\BibitemShut {NoStop}%
\bibitem [{\citenamefont {Gerosa}\ \emph {et~al.}(2019)\citenamefont {Gerosa},
  \citenamefont {Ma}, \citenamefont {Wong}, \citenamefont {Berti},
  \citenamefont {O’Shaughnessy}, \citenamefont {Chen},\ and\ \citenamefont
  {Belczynski}}]{gerosa2019multiband}%
  \BibitemOpen
  \bibfield  {author} {\bibinfo {author} {\bibfnamefont {D.}~\bibnamefont
  {Gerosa}}, \bibinfo {author} {\bibfnamefont {S.}~\bibnamefont {Ma}}, \bibinfo
  {author} {\bibfnamefont {K.~W.}\ \bibnamefont {Wong}}, \bibinfo {author}
  {\bibfnamefont {E.}~\bibnamefont {Berti}}, \bibinfo {author} {\bibfnamefont
  {R.}~\bibnamefont {O’Shaughnessy}}, \bibinfo {author} {\bibfnamefont
  {Y.}~\bibnamefont {Chen}},\ and\ \bibinfo {author} {\bibfnamefont
  {K.}~\bibnamefont {Belczynski}},\ }\bibfield  {title} {\bibinfo {title}
  {Multiband gravitational-wave event rates and stellar physics},\ }\href
  {https://doi.org/10.1103/PhysRevD.99.103004} {\bibfield  {journal} {\bibinfo
  {journal} {Physical Review D}\ }\textbf {\bibinfo {volume} {99}},\ \bibinfo
  {pages} {103004} (\bibinfo {year} {2019})}\BibitemShut {NoStop}%
\bibitem [{\citenamefont {Chamberlain}\ and\ \citenamefont
  {Yunes}(2017)}]{chamberlain2017theoretical}%
  \BibitemOpen
  \bibfield  {author} {\bibinfo {author} {\bibfnamefont {K.}~\bibnamefont
  {Chamberlain}}\ and\ \bibinfo {author} {\bibfnamefont {N.}~\bibnamefont
  {Yunes}},\ }\bibfield  {title} {\bibinfo {title} {Theoretical physics
  implications of gravitational wave observation with future detectors},\
  }\href {https://doi.org/10.1103/PhysRevD.96.084039} {\bibfield  {journal}
  {\bibinfo  {journal} {Physical Review D}\ }\textbf {\bibinfo {volume} {96}},\
  \bibinfo {pages} {084039} (\bibinfo {year} {2017})}\BibitemShut {NoStop}%
\bibitem [{\citenamefont {Gnocchi}\ \emph {et~al.}(2019)\citenamefont
  {Gnocchi}, \citenamefont {Maselli}, \citenamefont {Abdelsalhin},
  \citenamefont {Giacobbo},\ and\ \citenamefont
  {Mapelli}}]{gnocchi2019bounding}%
  \BibitemOpen
  \bibfield  {author} {\bibinfo {author} {\bibfnamefont {G.}~\bibnamefont
  {Gnocchi}}, \bibinfo {author} {\bibfnamefont {A.}~\bibnamefont {Maselli}},
  \bibinfo {author} {\bibfnamefont {T.}~\bibnamefont {Abdelsalhin}}, \bibinfo
  {author} {\bibfnamefont {N.}~\bibnamefont {Giacobbo}},\ and\ \bibinfo
  {author} {\bibfnamefont {M.}~\bibnamefont {Mapelli}},\ }\bibfield  {title}
  {\bibinfo {title} {Bounding alternative theories of gravity with multiband gw
  observations},\ }\href {https://doi.org/10.1103/PhysRevD.100.064024}
  {\bibfield  {journal} {\bibinfo  {journal} {Physical Review D}\ }\textbf
  {\bibinfo {volume} {100}},\ \bibinfo {pages} {064024} (\bibinfo {year}
  {2019})}\BibitemShut {NoStop}%
\bibitem [{\citenamefont {Tso}\ \emph {et~al.}(2019)\citenamefont {Tso},
  \citenamefont {Gerosa},\ and\ \citenamefont {Chen}}]{tso2019optimizing}%
  \BibitemOpen
  \bibfield  {author} {\bibinfo {author} {\bibfnamefont {R.}~\bibnamefont
  {Tso}}, \bibinfo {author} {\bibfnamefont {D.}~\bibnamefont {Gerosa}},\ and\
  \bibinfo {author} {\bibfnamefont {Y.}~\bibnamefont {Chen}},\ }\bibfield
  {title} {\bibinfo {title} {Optimizing ligo with lisa forewarnings to improve
  black-hole spectroscopy},\ }\href
  {https://doi.org/10.1103/PhysRevD.99.124043} {\bibfield  {journal} {\bibinfo
  {journal} {Physical Review D}\ }\textbf {\bibinfo {volume} {99}},\ \bibinfo
  {pages} {124043} (\bibinfo {year} {2019})}\BibitemShut {NoStop}%
\bibitem [{\citenamefont {Toubiana}\ \emph
  {et~al.}(2020{\natexlab{a}})\citenamefont {Toubiana}, \citenamefont {Marsat},
  \citenamefont {Babak}, \citenamefont {Barausse},\ and\ \citenamefont
  {Baker}}]{toubiana2020tests}%
  \BibitemOpen
  \bibfield  {author} {\bibinfo {author} {\bibfnamefont {A.}~\bibnamefont
  {Toubiana}}, \bibinfo {author} {\bibfnamefont {S.}~\bibnamefont {Marsat}},
  \bibinfo {author} {\bibfnamefont {S.}~\bibnamefont {Babak}}, \bibinfo
  {author} {\bibfnamefont {E.}~\bibnamefont {Barausse}},\ and\ \bibinfo
  {author} {\bibfnamefont {J.}~\bibnamefont {Baker}},\ }\bibfield  {title}
  {\bibinfo {title} {Tests of general relativity with stellar-mass black hole
  binaries observed by lisa},\ }\href
  {https://doi.org/10.1103/PhysRevD.101.104038} {\bibfield  {journal} {\bibinfo
   {journal} {Physical Review D}\ }\textbf {\bibinfo {volume} {101}},\ \bibinfo
  {pages} {104038} (\bibinfo {year} {2020}{\natexlab{a}})}\BibitemShut
  {NoStop}%
\bibitem [{\citenamefont {McGee}\ \emph {et~al.}(2020)\citenamefont {McGee},
  \citenamefont {Sesana},\ and\ \citenamefont {Vecchio}}]{mcgee2020linking}%
  \BibitemOpen
  \bibfield  {author} {\bibinfo {author} {\bibfnamefont {S.}~\bibnamefont
  {McGee}}, \bibinfo {author} {\bibfnamefont {A.}~\bibnamefont {Sesana}},\ and\
  \bibinfo {author} {\bibfnamefont {A.}~\bibnamefont {Vecchio}},\ }\bibfield
  {title} {\bibinfo {title} {Linking gravitational waves and x-ray phenomena
  with joint lisa and athena observations},\ }\href
  {https://doi.org/10.1038/s41550-019-0969-7} {\bibfield  {journal} {\bibinfo
  {journal} {Nature Astronomy}\ }\textbf {\bibinfo {volume} {4}},\ \bibinfo
  {pages} {26} (\bibinfo {year} {2020})}\BibitemShut {NoStop}%
\bibitem [{\citenamefont {{Caputo}}\ \emph {et~al.}(2020)\citenamefont
  {{Caputo}}, \citenamefont {{Sberna}}, \citenamefont {{Toubiana}},
  \citenamefont {{Babak}}, \citenamefont {{Barausse}} \emph
  {et~al.}}]{caputo2020gravitational}%
  \BibitemOpen
  \bibfield  {author} {\bibinfo {author} {\bibfnamefont {A.}~\bibnamefont
  {{Caputo}}}, \bibinfo {author} {\bibfnamefont {L.}~\bibnamefont {{Sberna}}},
  \bibinfo {author} {\bibfnamefont {A.}~\bibnamefont {{Toubiana}}}, \bibinfo
  {author} {\bibfnamefont {S.}~\bibnamefont {{Babak}}}, \bibinfo {author}
  {\bibfnamefont {E.}~\bibnamefont {{Barausse}}}, \emph {et~al.},\ }\bibfield
  {title} {\bibinfo {title} {{Gravitational-wave Detection and Parameter
  Estimation for Accreting Black-hole Binaries and Their Electromagnetic
  Counterpart}},\ }\href {https://doi.org/10.3847/1538-4357/ab7b66} {\bibfield
  {journal} {\bibinfo  {journal} {\apj}\ }\textbf {\bibinfo {volume} {892}},\
  \bibinfo {eid} {90} (\bibinfo {year} {2020})},\ \Eprint
  {https://arxiv.org/abs/2001.03620} {arXiv:2001.03620 [astro-ph.HE]}
  \BibitemShut {NoStop}%
\bibitem [{\citenamefont {Vitale}(2016)}]{vitale2016multiband}%
  \BibitemOpen
  \bibfield  {author} {\bibinfo {author} {\bibfnamefont {S.}~\bibnamefont
  {Vitale}},\ }\bibfield  {title} {\bibinfo {title} {Multiband
  gravitational-wave astronomy: parameter estimation and tests of general
  relativity with space-and ground-based detectors},\ }\href
  {https://doi.org/10.1103/PhysRevLett.117.051102} {\bibfield  {journal}
  {\bibinfo  {journal} {Physical Review Letters}\ }\textbf {\bibinfo {volume}
  {117}},\ \bibinfo {pages} {051102} (\bibinfo {year} {2016})}\BibitemShut
  {NoStop}%
\bibitem [{\citenamefont {Cornish}(2011)}]{cornish2011detection}%
  \BibitemOpen
  \bibfield  {author} {\bibinfo {author} {\bibfnamefont {N.~J.}\ \bibnamefont
  {Cornish}},\ }\bibfield  {title} {\bibinfo {title} {Detection strategies for
  extreme mass ratio inspirals},\ }\href
  {https://doi.org/10.1088/0264-9381/28/9/094016} {\bibfield  {journal}
  {\bibinfo  {journal} {Classical and Quantum Gravity}\ }\textbf {\bibinfo
  {volume} {28}},\ \bibinfo {pages} {094016} (\bibinfo {year}
  {2011})}\BibitemShut {NoStop}%
\bibitem [{\citenamefont {Chua}\ \emph {et~al.}(2017)\citenamefont {Chua},
  \citenamefont {Moore},\ and\ \citenamefont {Gair}}]{chua2017augmented}%
  \BibitemOpen
  \bibfield  {author} {\bibinfo {author} {\bibfnamefont {A.~J.}\ \bibnamefont
  {Chua}}, \bibinfo {author} {\bibfnamefont {C.~J.}\ \bibnamefont {Moore}},\
  and\ \bibinfo {author} {\bibfnamefont {J.~R.}\ \bibnamefont {Gair}},\
  }\bibfield  {title} {\bibinfo {title} {Augmented kludge waveforms for
  detecting extreme-mass-ratio inspirals},\ }\href
  {https://doi.org/10.1103/PhysRevD.96.044005} {\bibfield  {journal} {\bibinfo
  {journal} {Physical Review D}\ }\textbf {\bibinfo {volume} {96}},\ \bibinfo
  {pages} {044005} (\bibinfo {year} {2017})}\BibitemShut {NoStop}%
\bibitem [{\citenamefont {Babak}\ \emph {et~al.}(2017)\citenamefont {Babak},
  \citenamefont {Gair}, \citenamefont {Sesana}, \citenamefont {Barausse},
  \citenamefont {Sopuerta} \emph {et~al.}}]{babak2017science}%
  \BibitemOpen
  \bibfield  {author} {\bibinfo {author} {\bibfnamefont {S.}~\bibnamefont
  {Babak}}, \bibinfo {author} {\bibfnamefont {J.}~\bibnamefont {Gair}},
  \bibinfo {author} {\bibfnamefont {A.}~\bibnamefont {Sesana}}, \bibinfo
  {author} {\bibfnamefont {E.}~\bibnamefont {Barausse}}, \bibinfo {author}
  {\bibfnamefont {C.~F.}\ \bibnamefont {Sopuerta}}, \emph {et~al.},\ }\bibfield
   {title} {\bibinfo {title} {Science with the space-based interferometer lisa.
  v. extreme mass-ratio inspirals},\ }\href
  {https://doi.org/10.1103/PhysRevD.95.103012} {\bibfield  {journal} {\bibinfo
  {journal} {Physical Review D}\ }\textbf {\bibinfo {volume} {95}},\ \bibinfo
  {pages} {103012} (\bibinfo {year} {2017})}\BibitemShut {NoStop}%
\bibitem [{\citenamefont {Ewing}\ \emph {et~al.}(2021)\citenamefont {Ewing},
  \citenamefont {Sachdev}, \citenamefont {Borhanian},\ and\ \citenamefont
  {Sathyaprakash}}]{ewing2021archival}%
  \BibitemOpen
  \bibfield  {author} {\bibinfo {author} {\bibfnamefont {B.}~\bibnamefont
  {Ewing}}, \bibinfo {author} {\bibfnamefont {S.}~\bibnamefont {Sachdev}},
  \bibinfo {author} {\bibfnamefont {S.}~\bibnamefont {Borhanian}},\ and\
  \bibinfo {author} {\bibfnamefont {B.}~\bibnamefont {Sathyaprakash}},\
  }\bibfield  {title} {\bibinfo {title} {Archival searches for stellar-mass
  binary black holes in lisa data},\ }\href
  {https://doi.org/10.1103/PhysRevD.103.023025} {\bibfield  {journal} {\bibinfo
   {journal} {Physical Review D}\ }\textbf {\bibinfo {volume} {103}},\ \bibinfo
  {pages} {023025} (\bibinfo {year} {2021})}\BibitemShut {NoStop}%
\bibitem [{\citenamefont {{Wang}}\ \emph {et~al.}(2024)\citenamefont {{Wang}},
  \citenamefont {{Harry}}, \citenamefont {{Nitz}},\ and\ \citenamefont
  {{Hu}}}]{2024PhRvD.109f3029W}%
  \BibitemOpen
  \bibfield  {author} {\bibinfo {author} {\bibfnamefont {H.}~\bibnamefont
  {{Wang}}}, \bibinfo {author} {\bibfnamefont {I.}~\bibnamefont {{Harry}}},
  \bibinfo {author} {\bibfnamefont {A.}~\bibnamefont {{Nitz}}},\ and\ \bibinfo
  {author} {\bibfnamefont {Y.-M.}\ \bibnamefont {{Hu}}},\ }\bibfield  {title}
  {\bibinfo {title} {{Space-based gravitational wave observatories will be able
  to use eccentricity to unveil stellar-mass binary black hole formation}},\
  }\href {https://doi.org/10.1103/PhysRevD.109.063029} {\bibfield  {journal}
  {\bibinfo  {journal} {\prd}\ }\textbf {\bibinfo {volume} {109}},\ \bibinfo
  {eid} {063029} (\bibinfo {year} {2024})},\ \Eprint
  {https://arxiv.org/abs/2304.10340} {arXiv:2304.10340 [astro-ph.HE]}
  \BibitemShut {NoStop}%
\bibitem [{\citenamefont {Buscicchio}\ \emph {et~al.}(2021)\citenamefont
  {Buscicchio}, \citenamefont {Klein}, \citenamefont {Roebber}, \citenamefont
  {Moore}, \citenamefont {Gerosa}, \citenamefont {Finch},\ and\ \citenamefont
  {Vecchio}}]{buscicchio2021bayesian}%
  \BibitemOpen
  \bibfield  {author} {\bibinfo {author} {\bibfnamefont {R.}~\bibnamefont
  {Buscicchio}}, \bibinfo {author} {\bibfnamefont {A.}~\bibnamefont {Klein}},
  \bibinfo {author} {\bibfnamefont {E.}~\bibnamefont {Roebber}}, \bibinfo
  {author} {\bibfnamefont {C.~J.}\ \bibnamefont {Moore}}, \bibinfo {author}
  {\bibfnamefont {D.}~\bibnamefont {Gerosa}}, \bibinfo {author} {\bibfnamefont
  {E.}~\bibnamefont {Finch}},\ and\ \bibinfo {author} {\bibfnamefont
  {A.}~\bibnamefont {Vecchio}},\ }\bibfield  {title} {\bibinfo {title}
  {Bayesian parameter estimation of stellar-mass black-hole binaries with
  lisa},\ }\href {https://doi.org/10.1103/PhysRevD.104.044065} {\bibfield
  {journal} {\bibinfo  {journal} {Physical Review D}\ }\textbf {\bibinfo
  {volume} {104}},\ \bibinfo {pages} {044065} (\bibinfo {year}
  {2021})}\BibitemShut {NoStop}%
\bibitem [{\citenamefont {Toubiana}\ \emph
  {et~al.}(2020{\natexlab{b}})\citenamefont {Toubiana}, \citenamefont {Marsat},
  \citenamefont {Babak}, \citenamefont {Baker},\ and\ \citenamefont
  {Dal~Canton}}]{toubiana2020parameter}%
  \BibitemOpen
  \bibfield  {author} {\bibinfo {author} {\bibfnamefont {A.}~\bibnamefont
  {Toubiana}}, \bibinfo {author} {\bibfnamefont {S.}~\bibnamefont {Marsat}},
  \bibinfo {author} {\bibfnamefont {S.}~\bibnamefont {Babak}}, \bibinfo
  {author} {\bibfnamefont {J.}~\bibnamefont {Baker}},\ and\ \bibinfo {author}
  {\bibfnamefont {T.}~\bibnamefont {Dal~Canton}},\ }\bibfield  {title}
  {\bibinfo {title} {Parameter estimation of stellar-mass black hole binaries
  with lisa},\ }\href {https://doi.org/10.1103/PhysRevD.102.124037} {\bibfield
  {journal} {\bibinfo  {journal} {Physical Review D}\ }\textbf {\bibinfo
  {volume} {102}},\ \bibinfo {pages} {124037} (\bibinfo {year}
  {2020}{\natexlab{b}})}\BibitemShut {NoStop}%
\bibitem [{\citenamefont {{Lyu}}\ \emph {et~al.}(2023)\citenamefont {{Lyu}},
  \citenamefont {{Li}},\ and\ \citenamefont {{Hu}}}]{2023arXiv230712244L}%
  \BibitemOpen
  \bibfield  {author} {\bibinfo {author} {\bibfnamefont {X.}~\bibnamefont
  {{Lyu}}}, \bibinfo {author} {\bibfnamefont {E.-K.}\ \bibnamefont {{Li}}},\
  and\ \bibinfo {author} {\bibfnamefont {Y.-M.}\ \bibnamefont {{Hu}}},\
  }\bibfield  {title} {\bibinfo {title} {{Parameter Estimation of Stellar Mass
  Binary Black Holes under the Network of TianQin and LISA}},\ }\href
  {https://doi.org/10.48550/arXiv.2307.12244} {\bibfield  {journal} {\bibinfo
  {journal} {arXiv e-prints}\ ,\ \bibinfo {eid} {arXiv:2307.12244}} (\bibinfo
  {year} {2023})},\ \Eprint {https://arxiv.org/abs/2307.12244}
  {arXiv:2307.12244 [gr-qc]} \BibitemShut {NoStop}%
\bibitem [{\citenamefont {Kennedy}\ and\ \citenamefont
  {Eberhart}(1995)}]{kennedy1995particle}%
  \BibitemOpen
  \bibfield  {author} {\bibinfo {author} {\bibfnamefont {J.}~\bibnamefont
  {Kennedy}}\ and\ \bibinfo {author} {\bibfnamefont {R.}~\bibnamefont
  {Eberhart}},\ }\bibfield  {title} {\bibinfo {title} {Particle swarm
  optimization},\ }in\ \href@noop {} {\emph {\bibinfo {booktitle} {Proceedings
  of ICNN'95 - International Conference on Neural Networks}}},\ Vol.~\bibinfo
  {volume} {4}\ (\bibinfo {organization} {IEEE},\ \bibinfo {year} {1995})\ pp.\
  \bibinfo {pages} {1942--1948}\BibitemShut {NoStop}%
\bibitem [{\citenamefont {Shi}(1998)}]{shi1998modified}%
  \BibitemOpen
  \bibfield  {author} {\bibinfo {author} {\bibfnamefont {Y.}~\bibnamefont
  {Shi}},\ }\bibfield  {title} {\bibinfo {title} {A modified particle swarm
  optimizer},\ }in\ \href@noop {} {\emph {\bibinfo {booktitle} {Proc of IEEE
  Icec Conference}}}\ (\bibinfo {year} {1998})\BibitemShut {NoStop}%
\bibitem [{\citenamefont {Mohanty}(2018)}]{mohanty2018swarm}%
  \BibitemOpen
  \bibfield  {author} {\bibinfo {author} {\bibfnamefont {S.}~\bibnamefont
  {Mohanty}},\ }\href@noop {} {\emph {\bibinfo {title} {Swarm intelligence
  methods for statistical regression}}}\ (\bibinfo  {publisher} {CRC Press},\
  \bibinfo {year} {2018})\BibitemShut {NoStop}%
\bibitem [{\citenamefont {{Weerathunga}}\ and\ \citenamefont
  {{Mohanty}}(2017)}]{2017PhRvD..95l4030W}%
  \BibitemOpen
  \bibfield  {author} {\bibinfo {author} {\bibfnamefont {T.~S.}\ \bibnamefont
  {{Weerathunga}}}\ and\ \bibinfo {author} {\bibfnamefont {S.~D.}\ \bibnamefont
  {{Mohanty}}},\ }\bibfield  {title} {\bibinfo {title} {{Performance of
  particle swarm optimization on the fully-coherent all-sky search for
  gravitational waves from compact binary coalescences}},\ }\href
  {https://doi.org/10.1103/PhysRevD.95.124030} {\bibfield  {journal} {\bibinfo
  {journal} {\prd}\ }\textbf {\bibinfo {volume} {95}},\ \bibinfo {eid} {124030}
  (\bibinfo {year} {2017})},\ \Eprint {https://arxiv.org/abs/1703.01521}
  {arXiv:1703.01521 [gr-qc]} \BibitemShut {NoStop}%
\bibitem [{\citenamefont {{Normandin}}\ and\ \citenamefont
  {{Mohanty}}(2020)}]{2020PhRvD.101h2001N}%
  \BibitemOpen
  \bibfield  {author} {\bibinfo {author} {\bibfnamefont {M.~E.}\ \bibnamefont
  {{Normandin}}}\ and\ \bibinfo {author} {\bibfnamefont {S.~D.}\ \bibnamefont
  {{Mohanty}}},\ }\bibfield  {title} {\bibinfo {title} {{Towards a real-time
  fully-coherent all-sky search for gravitational waves from compact binary
  coalescences using particle swarm optimization}},\ }\href
  {https://doi.org/10.1103/PhysRevD.101.082001} {\bibfield  {journal} {\bibinfo
   {journal} {\prd}\ }\textbf {\bibinfo {volume} {101}},\ \bibinfo {eid}
  {082001} (\bibinfo {year} {2020})},\ \Eprint
  {https://arxiv.org/abs/2002.02150} {arXiv:2002.02150 [astro-ph.IM]}
  \BibitemShut {NoStop}%
\bibitem [{\citenamefont {{Zhang}}\ \emph {et~al.}(2021)\citenamefont
  {{Zhang}}, \citenamefont {{Mohanty}}, \citenamefont {{Zou}},\ and\
  \citenamefont {{Liu}}}]{2021PhRvD.104b4023Z}%
  \BibitemOpen
  \bibfield  {author} {\bibinfo {author} {\bibfnamefont {X.-H.}\ \bibnamefont
  {{Zhang}}}, \bibinfo {author} {\bibfnamefont {S.~D.}\ \bibnamefont
  {{Mohanty}}}, \bibinfo {author} {\bibfnamefont {X.-B.}\ \bibnamefont
  {{Zou}}},\ and\ \bibinfo {author} {\bibfnamefont {Y.-X.}\ \bibnamefont
  {{Liu}}},\ }\bibfield  {title} {\bibinfo {title} {{Resolving Galactic
  binaries in LISA data using particle swarm optimization and
  cross-validation}},\ }\href {https://doi.org/10.1103/PhysRevD.104.024023}
  {\bibfield  {journal} {\bibinfo  {journal} {\prd}\ }\textbf {\bibinfo
  {volume} {104}},\ \bibinfo {eid} {024023} (\bibinfo {year} {2021})},\ \Eprint
  {https://arxiv.org/abs/2103.09391} {arXiv:2103.09391 [gr-qc]} \BibitemShut
  {NoStop}%
\bibitem [{\citenamefont {{Wang}}\ \emph {et~al.}(2015)\citenamefont {{Wang}},
  \citenamefont {{Mohanty}},\ and\ \citenamefont
  {{Jenet}}}]{2015ApJ...815..125W}%
  \BibitemOpen
  \bibfield  {author} {\bibinfo {author} {\bibfnamefont {Y.}~\bibnamefont
  {{Wang}}}, \bibinfo {author} {\bibfnamefont {S.~D.}\ \bibnamefont
  {{Mohanty}}},\ and\ \bibinfo {author} {\bibfnamefont {F.~A.}\ \bibnamefont
  {{Jenet}}},\ }\bibfield  {title} {\bibinfo {title} {{Coherent Network
  Analysis for Continuous Gravitational Wave Signals in a Pulsar Timing Array:
  Pulsar Phases as Extrinsic Parameters}},\ }\href
  {https://doi.org/10.1088/0004-637X/815/2/125} {\bibfield  {journal} {\bibinfo
   {journal} {\apj}\ }\textbf {\bibinfo {volume} {815}},\ \bibinfo {eid} {125}
  (\bibinfo {year} {2015})},\ \Eprint {https://arxiv.org/abs/1506.01526}
  {arXiv:1506.01526 [astro-ph.IM]} \BibitemShut {NoStop}%
\bibitem [{\citenamefont {{Wang}}\ and\ \citenamefont
  {{Mohanty}}(2017)}]{2017PhRvL.118o1104W}%
  \BibitemOpen
  \bibfield  {author} {\bibinfo {author} {\bibfnamefont {Y.}~\bibnamefont
  {{Wang}}}\ and\ \bibinfo {author} {\bibfnamefont {S.~D.}\ \bibnamefont
  {{Mohanty}}},\ }\bibfield  {title} {\bibinfo {title} {{Pulsar Timing Array
  Based Search for Supermassive Black Hole Binaries in the Square Kilometer
  Array Era}},\ }\href {https://doi.org/10.1103/PhysRevLett.118.151104}
  {\bibfield  {journal} {\bibinfo  {journal} {\prl}\ }\textbf {\bibinfo
  {volume} {118}},\ \bibinfo {eid} {151104} (\bibinfo {year} {2017})},\ \Eprint
  {https://arxiv.org/abs/1611.09440} {arXiv:1611.09440 [astro-ph.IM]}
  \BibitemShut {NoStop}%
\bibitem [{\citenamefont {{Riles}}(2023)}]{riles2206searches}%
  \BibitemOpen
  \bibfield  {author} {\bibinfo {author} {\bibfnamefont {K.}~\bibnamefont
  {{Riles}}},\ }\bibfield  {title} {\bibinfo {title} {{Searches for
  continuous-wave gravitational radiation}},\ }\href
  {https://doi.org/10.1007/s41114-023-00044-3} {\bibfield  {journal} {\bibinfo
  {journal} {Living Reviews in Relativity}\ }\textbf {\bibinfo {volume} {26}},\
  \bibinfo {eid} {3} (\bibinfo {year} {2023})},\ \Eprint
  {https://arxiv.org/abs/2206.06447} {arXiv:2206.06447 [astro-ph.HE]}
  \BibitemShut {NoStop}%
\bibitem [{\citenamefont {Steltner}\ \emph {et~al.}(2021)\citenamefont
  {Steltner}, \citenamefont {Papa}, \citenamefont {Eggenstein}, \citenamefont
  {Allen}, \citenamefont {Dergachev}, \citenamefont {Prix}, \citenamefont
  {Machenschalk}, \citenamefont {Walsh}, \citenamefont {Zhu}, \citenamefont
  {Behnke} \emph {et~al.}}]{steltner2021einstein}%
  \BibitemOpen
  \bibfield  {author} {\bibinfo {author} {\bibfnamefont {B.}~\bibnamefont
  {Steltner}}, \bibinfo {author} {\bibfnamefont {M.~A.}\ \bibnamefont {Papa}},
  \bibinfo {author} {\bibfnamefont {H.-B.}\ \bibnamefont {Eggenstein}},
  \bibinfo {author} {\bibfnamefont {B.}~\bibnamefont {Allen}}, \bibinfo
  {author} {\bibfnamefont {V.}~\bibnamefont {Dergachev}}, \bibinfo {author}
  {\bibfnamefont {R.}~\bibnamefont {Prix}}, \bibinfo {author} {\bibfnamefont
  {B.}~\bibnamefont {Machenschalk}}, \bibinfo {author} {\bibfnamefont
  {S.}~\bibnamefont {Walsh}}, \bibinfo {author} {\bibfnamefont {S.~J.}\
  \bibnamefont {Zhu}}, \bibinfo {author} {\bibfnamefont {O.}~\bibnamefont
  {Behnke}}, \emph {et~al.},\ }\bibfield  {title} {\bibinfo {title} {Einstein@
  home all-sky search for continuous gravitational waves in ligo o2 public
  data},\ }\href {https://doi.org/10.3847/1538-4357/abc7c9} {\bibfield
  {journal} {\bibinfo  {journal} {The Astrophysical Journal}\ }\textbf
  {\bibinfo {volume} {909}},\ \bibinfo {pages} {79} (\bibinfo {year}
  {2021})}\BibitemShut {NoStop}%
\bibitem [{\citenamefont {Steltner}\ \emph {et~al.}(2023)\citenamefont
  {Steltner}, \citenamefont {Papa}, \citenamefont {Eggenstein}, \citenamefont
  {Prix}, \citenamefont {Bensch}, \citenamefont {Allen},\ and\ \citenamefont
  {Machenschalk}}]{steltner2023deep}%
  \BibitemOpen
  \bibfield  {author} {\bibinfo {author} {\bibfnamefont {B.}~\bibnamefont
  {Steltner}}, \bibinfo {author} {\bibfnamefont {M.}~\bibnamefont {Papa}},
  \bibinfo {author} {\bibfnamefont {H.-B.}\ \bibnamefont {Eggenstein}},
  \bibinfo {author} {\bibfnamefont {R.}~\bibnamefont {Prix}}, \bibinfo {author}
  {\bibfnamefont {M.}~\bibnamefont {Bensch}}, \bibinfo {author} {\bibfnamefont
  {B.}~\bibnamefont {Allen}},\ and\ \bibinfo {author} {\bibfnamefont
  {B.}~\bibnamefont {Machenschalk}},\ }\bibfield  {title} {\bibinfo {title}
  {Deep einstein@ home all-sky search for continuous gravitational waves in
  ligo o3 public data},\ }\href {https://doi.org/10.3847/1538-4357/abc7c9}
  {\bibfield  {journal} {\bibinfo  {journal} {The Astrophysical Journal}\
  }\textbf {\bibinfo {volume} {952}},\ \bibinfo {pages} {55} (\bibinfo {year}
  {2023})}\BibitemShut {NoStop}%
\bibitem [{\citenamefont {{Bandopadhyay}}\ and\ \citenamefont
  {{Moore}}(2023)}]{bandopadhyay2023lisa}%
  \BibitemOpen
  \bibfield  {author} {\bibinfo {author} {\bibfnamefont {D.}~\bibnamefont
  {{Bandopadhyay}}}\ and\ \bibinfo {author} {\bibfnamefont {C.~J.}\
  \bibnamefont {{Moore}}},\ }\bibfield  {title} {\bibinfo {title} {{LISA
  stellar-mass black hole searches with semicoherent and particle-swarm
  methods}},\ }\href {https://doi.org/10.1103/PhysRevD.108.084014} {\bibfield
  {journal} {\bibinfo  {journal} {\prd}\ }\textbf {\bibinfo {volume} {108}},\
  \bibinfo {eid} {084014} (\bibinfo {year} {2023})},\ \Eprint
  {https://arxiv.org/abs/2305.18048} {arXiv:2305.18048 [gr-qc]} \BibitemShut
  {NoStop}%
\bibitem [{\citenamefont {Jaranowski}\ \emph {et~al.}(1998)\citenamefont
  {Jaranowski}, \citenamefont {Krolak},\ and\ \citenamefont
  {Schutz}}]{jaranowski1998data}%
  \BibitemOpen
  \bibfield  {author} {\bibinfo {author} {\bibfnamefont {P.}~\bibnamefont
  {Jaranowski}}, \bibinfo {author} {\bibfnamefont {A.}~\bibnamefont {Krolak}},\
  and\ \bibinfo {author} {\bibfnamefont {B.~F.}\ \bibnamefont {Schutz}},\
  }\bibfield  {title} {\bibinfo {title} {Data analysis of gravitational-wave
  signals from spinning neutron stars: The signal and its detection},\ }\href
  {https://doi.org/10.1103/PhysRevD.58.063001} {\bibfield  {journal} {\bibinfo
  {journal} {Physical Review D}\ }\textbf {\bibinfo {volume} {58}},\ \bibinfo
  {pages} {063001} (\bibinfo {year} {1998})}\BibitemShut {NoStop}%
\bibitem [{\citenamefont {{Armstrong}}\ \emph {et~al.}(1999)\citenamefont
  {{Armstrong}}, \citenamefont {{Estabrook}},\ and\ \citenamefont
  {{Tinto}}}]{1999ApJ...527..814A}%
  \BibitemOpen
  \bibfield  {author} {\bibinfo {author} {\bibfnamefont {J.~W.}\ \bibnamefont
  {{Armstrong}}}, \bibinfo {author} {\bibfnamefont {F.~B.}\ \bibnamefont
  {{Estabrook}}},\ and\ \bibinfo {author} {\bibfnamefont {M.}~\bibnamefont
  {{Tinto}}},\ }\bibfield  {title} {\bibinfo {title} {{Time-Delay
  Interferometry for Space-based Gravitational Wave Searches}},\ }\href
  {https://doi.org/10.1086/308110} {\bibfield  {journal} {\bibinfo  {journal}
  {\apj}\ }\textbf {\bibinfo {volume} {527}},\ \bibinfo {pages} {814} (\bibinfo
  {year} {1999})}\BibitemShut {NoStop}%
\bibitem [{\citenamefont {{Tinto}}\ and\ \citenamefont
  {{Dhurandhar}}(2005)}]{2005LRR.....8....4T}%
  \BibitemOpen
  \bibfield  {author} {\bibinfo {author} {\bibfnamefont {M.}~\bibnamefont
  {{Tinto}}}\ and\ \bibinfo {author} {\bibfnamefont {S.~V.}\ \bibnamefont
  {{Dhurandhar}}},\ }\bibfield  {title} {\bibinfo {title} {{Time-Delay
  Interferometry}},\ }\href {https://doi.org/10.12942/lrr-2005-4} {\bibfield
  {journal} {\bibinfo  {journal} {Living Reviews in Relativity}\ }\textbf
  {\bibinfo {volume} {8}},\ \bibinfo {eid} {4} (\bibinfo {year}
  {2005})}\BibitemShut {NoStop}%
\bibitem [{\citenamefont {Prince}\ \emph {et~al.}(2002)\citenamefont {Prince},
  \citenamefont {Tinto}, \citenamefont {Larson},\ and\ \citenamefont
  {Armstrong}}]{PhysRevD.66.122002}%
  \BibitemOpen
  \bibfield  {author} {\bibinfo {author} {\bibfnamefont {T.~A.}\ \bibnamefont
  {Prince}}, \bibinfo {author} {\bibfnamefont {M.}~\bibnamefont {Tinto}},
  \bibinfo {author} {\bibfnamefont {S.~L.}\ \bibnamefont {Larson}},\ and\
  \bibinfo {author} {\bibfnamefont {J.~W.}\ \bibnamefont {Armstrong}},\
  }\bibfield  {title} {\bibinfo {title} {Lisa optimal sensitivity},\ }\href
  {https://doi.org/10.1103/PhysRevD.66.122002} {\bibfield  {journal} {\bibinfo
  {journal} {Phys. Rev. D}\ }\textbf {\bibinfo {volume} {66}},\ \bibinfo
  {pages} {122002} (\bibinfo {year} {2002})}\BibitemShut {NoStop}%
\bibitem [{\citenamefont {Kay}(1993)}]{10.5555/151045}%
  \BibitemOpen
  \bibfield  {author} {\bibinfo {author} {\bibfnamefont {S.~M.}\ \bibnamefont
  {Kay}},\ }\href@noop {} {\emph {\bibinfo {title} {{Fundamentals of
  statistical signal processing: estimation theory}}}}\ (\bibinfo  {publisher}
  {Prentice-Hall, Inc.},\ \bibinfo {address} {USA},\ \bibinfo {year}
  {1993})\BibitemShut {NoStop}%
\bibitem [{\citenamefont {{Finn}}(1992)}]{1992PhRvD..46.5236F}%
  \BibitemOpen
  \bibfield  {author} {\bibinfo {author} {\bibfnamefont {L.~S.}\ \bibnamefont
  {{Finn}}},\ }\bibfield  {title} {\bibinfo {title} {{Detection, measurement,
  and gravitational radiation}},\ }\href
  {https://doi.org/10.1103/PhysRevD.46.5236} {\bibfield  {journal} {\bibinfo
  {journal} {\prd}\ }\textbf {\bibinfo {volume} {46}},\ \bibinfo {pages} {5236}
  (\bibinfo {year} {1992})},\ \Eprint {https://arxiv.org/abs/gr-qc/9209010}
  {arXiv:gr-qc/9209010 [gr-qc]} \BibitemShut {NoStop}%
\bibitem [{\citenamefont {Crowder}\ and\ \citenamefont
  {Cornish}(2007)}]{crowder2007solution}%
  \BibitemOpen
  \bibfield  {author} {\bibinfo {author} {\bibfnamefont {J.}~\bibnamefont
  {Crowder}}\ and\ \bibinfo {author} {\bibfnamefont {N.~J.}\ \bibnamefont
  {Cornish}},\ }\bibfield  {title} {\bibinfo {title} {Solution to the galactic
  foreground problem for lisa},\ }\href
  {https://doi.org/10.1103/PhysRevD.75.043008} {\bibfield  {journal} {\bibinfo
  {journal} {Physical Review D}\ }\textbf {\bibinfo {volume} {75}},\ \bibinfo
  {pages} {043008} (\bibinfo {year} {2007})}\BibitemShut {NoStop}%
\bibitem [{\citenamefont {{Brady}}\ and\ \citenamefont
  {{Creighton}}(2000)}]{2000PhRvD..61h2001B}%
  \BibitemOpen
  \bibfield  {author} {\bibinfo {author} {\bibfnamefont {P.~R.}\ \bibnamefont
  {{Brady}}}\ and\ \bibinfo {author} {\bibfnamefont {T.}~\bibnamefont
  {{Creighton}}},\ }\bibfield  {title} {\bibinfo {title} {{Searching for
  periodic sources with LIGO. II. Hierarchical searches}},\ }\href
  {https://doi.org/10.1103/PhysRevD.61.082001} {\bibfield  {journal} {\bibinfo
  {journal} {\prd}\ }\textbf {\bibinfo {volume} {61}},\ \bibinfo {eid} {082001}
  (\bibinfo {year} {2000})},\ \Eprint {https://arxiv.org/abs/gr-qc/9812014}
  {arXiv:gr-qc/9812014 [gr-qc]} \BibitemShut {NoStop}%
\bibitem [{\citenamefont {{Cutler}}\ \emph {et~al.}(2005)\citenamefont
  {{Cutler}}, \citenamefont {{Gholami}},\ and\ \citenamefont
  {{Krishnan}}}]{2005PhRvD..72d2004C}%
  \BibitemOpen
  \bibfield  {author} {\bibinfo {author} {\bibfnamefont {C.}~\bibnamefont
  {{Cutler}}}, \bibinfo {author} {\bibfnamefont {I.}~\bibnamefont
  {{Gholami}}},\ and\ \bibinfo {author} {\bibfnamefont {B.}~\bibnamefont
  {{Krishnan}}},\ }\bibfield  {title} {\bibinfo {title} {{Improved stack-slide
  searches for gravitational-wave pulsars}},\ }\href
  {https://doi.org/10.1103/PhysRevD.72.042004} {\bibfield  {journal} {\bibinfo
  {journal} {\prd}\ }\textbf {\bibinfo {volume} {72}},\ \bibinfo {eid} {042004}
  (\bibinfo {year} {2005})},\ \Eprint {https://arxiv.org/abs/gr-qc/0505082}
  {arXiv:gr-qc/0505082 [gr-qc]} \BibitemShut {NoStop}%
\bibitem [{\citenamefont {{Krishnan}}\ \emph {et~al.}(2004)\citenamefont
  {{Krishnan}}, \citenamefont {{Sintes}}, \citenamefont {{Papa}}, \citenamefont
  {{Schutz}}, \citenamefont {{Frasca}},\ and\ \citenamefont
  {{Palomba}}}]{2004PhRvD..70h2001K}%
  \BibitemOpen
  \bibfield  {author} {\bibinfo {author} {\bibfnamefont {B.}~\bibnamefont
  {{Krishnan}}}, \bibinfo {author} {\bibfnamefont {A.~M.}\ \bibnamefont
  {{Sintes}}}, \bibinfo {author} {\bibfnamefont {M.~A.}\ \bibnamefont
  {{Papa}}}, \bibinfo {author} {\bibfnamefont {B.~F.}\ \bibnamefont
  {{Schutz}}}, \bibinfo {author} {\bibfnamefont {S.}~\bibnamefont {{Frasca}}},\
  and\ \bibinfo {author} {\bibfnamefont {C.}~\bibnamefont {{Palomba}}},\
  }\bibfield  {title} {\bibinfo {title} {{Hough transform search for continuous
  gravitational waves}},\ }\href {https://doi.org/10.1103/PhysRevD.70.082001}
  {\bibfield  {journal} {\bibinfo  {journal} {\prd}\ }\textbf {\bibinfo
  {volume} {70}},\ \bibinfo {eid} {082001} (\bibinfo {year} {2004})},\ \Eprint
  {https://arxiv.org/abs/gr-qc/0407001} {arXiv:gr-qc/0407001 [gr-qc]}
  \BibitemShut {NoStop}%
\bibitem [{\citenamefont {Antonucci}\ \emph {et~al.}(2008)\citenamefont
  {Antonucci}, \citenamefont {Astone}, \citenamefont {D'Antonio}, \citenamefont
  {Frasca},\ and\ \citenamefont {Palomba}}]{Antonucci_2008}%
  \BibitemOpen
  \bibfield  {author} {\bibinfo {author} {\bibfnamefont {F.}~\bibnamefont
  {Antonucci}}, \bibinfo {author} {\bibfnamefont {P.}~\bibnamefont {Astone}},
  \bibinfo {author} {\bibfnamefont {S.}~\bibnamefont {D'Antonio}}, \bibinfo
  {author} {\bibfnamefont {S.}~\bibnamefont {Frasca}},\ and\ \bibinfo {author}
  {\bibfnamefont {C.}~\bibnamefont {Palomba}},\ }\bibfield  {title} {\bibinfo
  {title} {Detection of periodic gravitational wave sources by hough transform
  in the f versus plane},\ }\href
  {https://doi.org/10.1088/0264-9381/25/18/184015} {\bibfield  {journal}
  {\bibinfo  {journal} {Classical and Quantum Gravity}\ }\textbf {\bibinfo
  {volume} {25}},\ \bibinfo {pages} {184015} (\bibinfo {year}
  {2008})}\BibitemShut {NoStop}%
\bibitem [{\citenamefont {{Dhurandhar}}\ \emph {et~al.}(2008)\citenamefont
  {{Dhurandhar}}, \citenamefont {{Krishnan}}, \citenamefont {{Mukhopadhyay}},\
  and\ \citenamefont {{Whelan}}}]{2008PhRvD..77h2001D}%
  \BibitemOpen
  \bibfield  {author} {\bibinfo {author} {\bibfnamefont {S.}~\bibnamefont
  {{Dhurandhar}}}, \bibinfo {author} {\bibfnamefont {B.}~\bibnamefont
  {{Krishnan}}}, \bibinfo {author} {\bibfnamefont {H.}~\bibnamefont
  {{Mukhopadhyay}}},\ and\ \bibinfo {author} {\bibfnamefont {J.~T.}\
  \bibnamefont {{Whelan}}},\ }\bibfield  {title} {\bibinfo {title}
  {{Cross-correlation search for periodic gravitational waves}},\ }\href
  {https://doi.org/10.1103/PhysRevD.77.082001} {\bibfield  {journal} {\bibinfo
  {journal} {\prd}\ }\textbf {\bibinfo {volume} {77}},\ \bibinfo {eid} {082001}
  (\bibinfo {year} {2008})},\ \Eprint {https://arxiv.org/abs/0712.1578}
  {arXiv:0712.1578 [gr-qc]} \BibitemShut {NoStop}%
\bibitem [{\citenamefont {Bratton}\ and\ \citenamefont
  {Kennedy}(2007)}]{4223164}%
  \BibitemOpen
  \bibfield  {author} {\bibinfo {author} {\bibfnamefont {D.}~\bibnamefont
  {Bratton}}\ and\ \bibinfo {author} {\bibfnamefont {J.}~\bibnamefont
  {Kennedy}},\ }\bibfield  {title} {\bibinfo {title} {Defining a standard for
  particle swarm optimization},\ }in\ \href
  {https://doi.org/10.1109/SIS.2007.368035} {\emph {\bibinfo {booktitle} {2007
  IEEE Swarm Intelligence Symposium}}}\ (\bibinfo {year} {2007})\ pp.\ \bibinfo
  {pages} {120--127}\BibitemShut {NoStop}%
\bibitem [{\citenamefont {Engelbercht}(2005)}]{engelbercht2005fundamentals}%
  \BibitemOpen
  \bibfield  {author} {\bibinfo {author} {\bibfnamefont {A.}~\bibnamefont
  {Engelbercht}},\ }\bibfield  {title} {\bibinfo {title} {Fundamentals of
  computational swarm intelligence},\ }\href@noop {} {\bibfield  {journal}
  {\bibinfo  {journal} {University of Pretoria, South Africa}\ } (\bibinfo
  {year} {2005})}\BibitemShut {NoStop}%
\bibitem [{\citenamefont {{Vallisneri}}(2005)}]{2005PhRvD..71b2001V}%
  \BibitemOpen
  \bibfield  {author} {\bibinfo {author} {\bibfnamefont {M.}~\bibnamefont
  {{Vallisneri}}},\ }\bibfield  {title} {\bibinfo {title} {{Synthetic LISA:
  Simulating time delay interferometry in a model LISA}},\ }\href
  {https://doi.org/10.1103/PhysRevD.71.022001} {\bibfield  {journal} {\bibinfo
  {journal} {\prd}\ }\textbf {\bibinfo {volume} {71}},\ \bibinfo {eid} {022001}
  (\bibinfo {year} {2005})},\ \Eprint {https://arxiv.org/abs/gr-qc/0407102}
  {arXiv:gr-qc/0407102 [gr-qc]} \BibitemShut {NoStop}%
\bibitem [{\citenamefont {{Marsat}}\ and\ \citenamefont
  {{Baker}}(2018)}]{marsat2018fourier}%
  \BibitemOpen
  \bibfield  {author} {\bibinfo {author} {\bibfnamefont {S.}~\bibnamefont
  {{Marsat}}}\ and\ \bibinfo {author} {\bibfnamefont {J.~G.}\ \bibnamefont
  {{Baker}}},\ }\bibfield  {title} {\bibinfo {title} {{Fourier-domain
  modulations and delays of gravitational-wave signals}},\ }\href
  {https://doi.org/10.48550/arXiv.1806.10734} {\bibfield  {journal} {\bibinfo
  {journal} {arXiv e-prints}\ ,\ \bibinfo {eid} {arXiv:1806.10734}} (\bibinfo
  {year} {2018})},\ \Eprint {https://arxiv.org/abs/1806.10734}
  {arXiv:1806.10734 [gr-qc]} \BibitemShut {NoStop}%
\bibitem [{\citenamefont {{Cornish}}\ and\ \citenamefont
  {{Rubbo}}(2003)}]{2003PhRvD..67b2001C}%
  \BibitemOpen
  \bibfield  {author} {\bibinfo {author} {\bibfnamefont {N.~J.}\ \bibnamefont
  {{Cornish}}}\ and\ \bibinfo {author} {\bibfnamefont {L.~J.}\ \bibnamefont
  {{Rubbo}}},\ }\bibfield  {title} {\bibinfo {title} {{LISA response
  function}},\ }\href {https://doi.org/10.1103/PhysRevD.67.022001} {\bibfield
  {journal} {\bibinfo  {journal} {\prd}\ }\textbf {\bibinfo {volume} {67}},\
  \bibinfo {eid} {022001} (\bibinfo {year} {2003})}\BibitemShut {NoStop}%
\bibitem [{\citenamefont {{Zhou}}\ \emph {et~al.}(2021)\citenamefont {{Zhou}},
  \citenamefont {{Hu}}, \citenamefont {{Ye}}, \citenamefont {{Hu}},
  \citenamefont {{Zhu}}, \citenamefont {{Zhang}}, \citenamefont {{Su}},\ and\
  \citenamefont {{Wang}}}]{2021PhRvD.103j3026Z}%
  \BibitemOpen
  \bibfield  {author} {\bibinfo {author} {\bibfnamefont {M.-Y.}\ \bibnamefont
  {{Zhou}}}, \bibinfo {author} {\bibfnamefont {X.-C.}\ \bibnamefont {{Hu}}},
  \bibinfo {author} {\bibfnamefont {B.}~\bibnamefont {{Ye}}}, \bibinfo {author}
  {\bibfnamefont {S.}~\bibnamefont {{Hu}}}, \bibinfo {author} {\bibfnamefont
  {D.-D.}\ \bibnamefont {{Zhu}}}, \bibinfo {author} {\bibfnamefont
  {X.}~\bibnamefont {{Zhang}}}, \bibinfo {author} {\bibfnamefont
  {W.}~\bibnamefont {{Su}}},\ and\ \bibinfo {author} {\bibfnamefont
  {Y.}~\bibnamefont {{Wang}}},\ }\bibfield  {title} {\bibinfo {title} {{Orbital
  effects on time delay interferometry for TianQin}},\ }\href
  {https://doi.org/10.1103/PhysRevD.103.103026} {\bibfield  {journal} {\bibinfo
   {journal} {\prd}\ }\textbf {\bibinfo {volume} {103}},\ \bibinfo {eid}
  {103026} (\bibinfo {year} {2021})},\ \Eprint
  {https://arxiv.org/abs/2102.10291} {arXiv:2102.10291 [astro-ph.IM]}
  \BibitemShut {NoStop}%
\bibitem [{\citenamefont {Rubbo}\ \emph {et~al.}(2004)\citenamefont {Rubbo},
  \citenamefont {Cornish},\ and\ \citenamefont {Poujade}}]{PhysRevD.69.082003}%
  \BibitemOpen
  \bibfield  {author} {\bibinfo {author} {\bibfnamefont {L.~J.}\ \bibnamefont
  {Rubbo}}, \bibinfo {author} {\bibfnamefont {N.~J.}\ \bibnamefont {Cornish}},\
  and\ \bibinfo {author} {\bibfnamefont {O.}~\bibnamefont {Poujade}},\
  }\bibfield  {title} {\bibinfo {title} {Forward modeling of space-borne
  gravitational wave detectors},\ }\href
  {https://doi.org/10.1103/PhysRevD.69.082003} {\bibfield  {journal} {\bibinfo
  {journal} {Phys. Rev. D}\ }\textbf {\bibinfo {volume} {69}},\ \bibinfo
  {pages} {082003} (\bibinfo {year} {2004})}\BibitemShut {NoStop}%
\bibitem [{\citenamefont {{Estabrook}}\ and\ \citenamefont
  {{Wahlquist}}(1975)}]{1975GReGr...6..439E}%
  \BibitemOpen
  \bibfield  {author} {\bibinfo {author} {\bibfnamefont {F.~B.}\ \bibnamefont
  {{Estabrook}}}\ and\ \bibinfo {author} {\bibfnamefont {H.~D.}\ \bibnamefont
  {{Wahlquist}}},\ }\bibfield  {title} {\bibinfo {title} {{Response of Doppler
  spacecraft tracking to gravitational radiation.}},\ }\href
  {https://doi.org/10.1007/BF00762449} {\bibfield  {journal} {\bibinfo
  {journal} {General Relativity and Gravitation}\ }\textbf {\bibinfo {volume}
  {6}},\ \bibinfo {pages} {439} (\bibinfo {year} {1975})}\BibitemShut {NoStop}%
\bibitem [{\citenamefont {{Wang}}\ and\ \citenamefont
  {{Hu}}(2023)}]{2023CoTPh..75g5402W}%
  \BibitemOpen
  \bibfield  {author} {\bibinfo {author} {\bibfnamefont {R.}~\bibnamefont
  {{Wang}}}\ and\ \bibinfo {author} {\bibfnamefont {B.}~\bibnamefont {{Hu}}},\
  }\bibfield  {title} {\bibinfo {title} {{LitePIG: a lite parameter inference
  system for the gravitational wave in the millihertz band}},\ }\href
  {https://doi.org/10.1088/1572-9494/acccdd} {\bibfield  {journal} {\bibinfo
  {journal} {Communications in Theoretical Physics}\ }\textbf {\bibinfo
  {volume} {75}},\ \bibinfo {eid} {075402} (\bibinfo {year} {2023})},\ \Eprint
  {https://arxiv.org/abs/2208.13351} {arXiv:2208.13351 [astro-ph.IM]}
  \BibitemShut {NoStop}%
\bibitem [{\citenamefont {Kumar}\ \emph {et~al.}(2019)\citenamefont {Kumar},
  \citenamefont {Blackman}, \citenamefont {Field}, \citenamefont {Scheel},
  \citenamefont {Galley}, \citenamefont {Boyle}, \citenamefont {Kidder},
  \citenamefont {Pfeiffer}, \citenamefont {Szilagyi},\ and\ \citenamefont
  {Teukolsky}}]{PhysRevD.99.124005}%
  \BibitemOpen
  \bibfield  {author} {\bibinfo {author} {\bibfnamefont {P.}~\bibnamefont
  {Kumar}}, \bibinfo {author} {\bibfnamefont {J.}~\bibnamefont {Blackman}},
  \bibinfo {author} {\bibfnamefont {S.~E.}\ \bibnamefont {Field}}, \bibinfo
  {author} {\bibfnamefont {M.}~\bibnamefont {Scheel}}, \bibinfo {author}
  {\bibfnamefont {C.~R.}\ \bibnamefont {Galley}}, \bibinfo {author}
  {\bibfnamefont {M.}~\bibnamefont {Boyle}}, \bibinfo {author} {\bibfnamefont
  {L.~E.}\ \bibnamefont {Kidder}}, \bibinfo {author} {\bibfnamefont {H.~P.}\
  \bibnamefont {Pfeiffer}}, \bibinfo {author} {\bibfnamefont {B.}~\bibnamefont
  {Szilagyi}},\ and\ \bibinfo {author} {\bibfnamefont {S.~A.}\ \bibnamefont
  {Teukolsky}},\ }\bibfield  {title} {\bibinfo {title} {Constraining the
  parameters of gw150914 and gw170104 with numerical relativity surrogates},\
  }\href {https://doi.org/10.1103/PhysRevD.99.124005} {\bibfield  {journal}
  {\bibinfo  {journal} {Phys. Rev. D}\ }\textbf {\bibinfo {volume} {99}},\
  \bibinfo {pages} {124005} (\bibinfo {year} {2019})}\BibitemShut {NoStop}%
\bibitem [{\citenamefont {Gong}\ \emph {et~al.}(2023)\citenamefont {Gong},
  \citenamefont {Cao}, \citenamefont {Zhao},\ and\ \citenamefont
  {Shao}}]{PhysRevD.108.064046}%
  \BibitemOpen
  \bibfield  {author} {\bibinfo {author} {\bibfnamefont {Y.}~\bibnamefont
  {Gong}}, \bibinfo {author} {\bibfnamefont {Z.}~\bibnamefont {Cao}}, \bibinfo
  {author} {\bibfnamefont {J.}~\bibnamefont {Zhao}},\ and\ \bibinfo {author}
  {\bibfnamefont {L.}~\bibnamefont {Shao}},\ }\bibfield  {title} {\bibinfo
  {title} {Including higher harmonics in gravitational-wave parameter
  estimation and cosmological implications for lisa},\ }\href
  {https://doi.org/10.1103/PhysRevD.108.064046} {\bibfield  {journal} {\bibinfo
   {journal} {Phys. Rev. D}\ }\textbf {\bibinfo {volume} {108}},\ \bibinfo
  {pages} {064046} (\bibinfo {year} {2023})}\BibitemShut {NoStop}%
\bibitem [{\citenamefont {Prasad}\ and\ \citenamefont
  {Souradeep}(2012)}]{PhysRevD.85.123008}%
  \BibitemOpen
  \bibfield  {author} {\bibinfo {author} {\bibfnamefont {J.}~\bibnamefont
  {Prasad}}\ and\ \bibinfo {author} {\bibfnamefont {T.}~\bibnamefont
  {Souradeep}},\ }\bibfield  {title} {\bibinfo {title} {Cosmological parameter
  estimation using particle swarm optimization},\ }\href
  {https://doi.org/10.1103/PhysRevD.85.123008} {\bibfield  {journal} {\bibinfo
  {journal} {Phys. Rev. D}\ }\textbf {\bibinfo {volume} {85}},\ \bibinfo
  {pages} {123008} (\bibinfo {year} {2012})}\BibitemShut {NoStop}%
\bibitem [{\citenamefont {{Feng}}\ \emph {et~al.}(2019)\citenamefont {{Feng}},
  \citenamefont {{Wang}}, \citenamefont {{Hu}}, \citenamefont {{Hu}},\ and\
  \citenamefont {{Wang}}}]{2019PhRvD..99l3002F}%
  \BibitemOpen
  \bibfield  {author} {\bibinfo {author} {\bibfnamefont {W.-F.}\ \bibnamefont
  {{Feng}}}, \bibinfo {author} {\bibfnamefont {H.-T.}\ \bibnamefont {{Wang}}},
  \bibinfo {author} {\bibfnamefont {X.-C.}\ \bibnamefont {{Hu}}}, \bibinfo
  {author} {\bibfnamefont {Y.-M.}\ \bibnamefont {{Hu}}},\ and\ \bibinfo
  {author} {\bibfnamefont {Y.}~\bibnamefont {{Wang}}},\ }\bibfield  {title}
  {\bibinfo {title} {{Preliminary study on parameter estimation accuracy of
  supermassive black hole binary inspirals for TianQin}},\ }\href
  {https://doi.org/10.1103/PhysRevD.99.123002} {\bibfield  {journal} {\bibinfo
  {journal} {\prd}\ }\textbf {\bibinfo {volume} {99}},\ \bibinfo {eid} {123002}
  (\bibinfo {year} {2019})},\ \Eprint {https://arxiv.org/abs/1901.02159}
  {arXiv:1901.02159 [astro-ph.IM]} \BibitemShut {NoStop}%
\bibitem [{\citenamefont {Cutler}(1998)}]{PhysRevD.57.7089}%
  \BibitemOpen
  \bibfield  {author} {\bibinfo {author} {\bibfnamefont {C.}~\bibnamefont
  {Cutler}},\ }\bibfield  {title} {\bibinfo {title} {Angular resolution of the
  lisa gravitational wave detector},\ }\href
  {https://doi.org/10.1103/PhysRevD.57.7089} {\bibfield  {journal} {\bibinfo
  {journal} {Phys. Rev. D}\ }\textbf {\bibinfo {volume} {57}},\ \bibinfo
  {pages} {7089} (\bibinfo {year} {1998})}\BibitemShut {NoStop}%
\bibitem [{\citenamefont {Kr\'olak}\ \emph {et~al.}(1995)\citenamefont
  {Kr\'olak}, \citenamefont {Kokkotas},\ and\ \citenamefont
  {Sch\"afer}}]{PhysRevD.52.2089}%
  \BibitemOpen
  \bibfield  {author} {\bibinfo {author} {\bibfnamefont {A.}~\bibnamefont
  {Kr\'olak}}, \bibinfo {author} {\bibfnamefont {K.~D.}\ \bibnamefont
  {Kokkotas}},\ and\ \bibinfo {author} {\bibfnamefont {G.}~\bibnamefont
  {Sch\"afer}},\ }\bibfield  {title} {\bibinfo {title} {Estimation of the
  post-newtonian parameters in the gravitational-wave emission of a coalescing
  binary},\ }\href {https://doi.org/10.1103/PhysRevD.52.2089} {\bibfield
  {journal} {\bibinfo  {journal} {Phys. Rev. D}\ }\textbf {\bibinfo {volume}
  {52}},\ \bibinfo {pages} {2089} (\bibinfo {year} {1995})}\BibitemShut
  {NoStop}%
\bibitem [{\citenamefont {Buonanno}\ \emph {et~al.}(2009)\citenamefont
  {Buonanno}, \citenamefont {Iyer}, \citenamefont {Ochsner}, \citenamefont
  {Pan},\ and\ \citenamefont {Sathyaprakash}}]{PhysRevD.80.084043}%
  \BibitemOpen
  \bibfield  {author} {\bibinfo {author} {\bibfnamefont {A.}~\bibnamefont
  {Buonanno}}, \bibinfo {author} {\bibfnamefont {B.~R.}\ \bibnamefont {Iyer}},
  \bibinfo {author} {\bibfnamefont {E.}~\bibnamefont {Ochsner}}, \bibinfo
  {author} {\bibfnamefont {Y.}~\bibnamefont {Pan}},\ and\ \bibinfo {author}
  {\bibfnamefont {B.~S.}\ \bibnamefont {Sathyaprakash}},\ }\bibfield  {title}
  {\bibinfo {title} {Comparison of post-newtonian templates for compact binary
  inspiral signals in gravitational-wave detectors},\ }\href
  {https://doi.org/10.1103/PhysRevD.80.084043} {\bibfield  {journal} {\bibinfo
  {journal} {Phys. Rev. D}\ }\textbf {\bibinfo {volume} {80}},\ \bibinfo
  {pages} {084043} (\bibinfo {year} {2009})}\BibitemShut {NoStop}%
\bibitem [{\citenamefont {{Crowder}}\ and\ \citenamefont
  {{Cornish}}(2007)}]{2007PhRvD..75d3008C}%
  \BibitemOpen
  \bibfield  {author} {\bibinfo {author} {\bibfnamefont {J.}~\bibnamefont
  {{Crowder}}}\ and\ \bibinfo {author} {\bibfnamefont {N.~J.}\ \bibnamefont
  {{Cornish}}},\ }\bibfield  {title} {\bibinfo {title} {{Solution to the
  galactic foreground problem for LISA}},\ }\href
  {https://doi.org/10.1103/PhysRevD.75.043008} {\bibfield  {journal} {\bibinfo
  {journal} {\prd}\ }\textbf {\bibinfo {volume} {75}},\ \bibinfo {eid} {043008}
  (\bibinfo {year} {2007})},\ \Eprint {https://arxiv.org/abs/astro-ph/0611546}
  {arXiv:astro-ph/0611546 [astro-ph]} \BibitemShut {NoStop}%
\end{thebibliography}

%

\end{document}